\documentclass[5p,times,twocolumn]{elsarticle}
\usepackage{lineno,hyperref}
\usepackage{subcaption}
\usepackage{tikz}
\usepackage{pgfplotstable}
\usepackage{pgfplots}
\usepackage{booktabs}
\usepackage{multirow}
\usepackage{float}
\usepackage{textcomp}
\usepackage{amsmath,amsfonts,amssymb,bbm}	
\usepackage{multicol}
\usepackage{blindtext}
\usepackage{siunitx}

\usepackage[section]{placeins}

\modulolinenumbers[5]

\biboptions{authoryear}

\begin{document}

\begin{frontmatter}

\title{Liver Segmentation using Turbolift Learning for CT and Cone-beam C-arm Perfusion Imaging}

\author[1,5]{Hana Haselji{\'c}\corref{equalcontribution}}
\author[2,3,4,5,6]{Soumick Chatterjee\corref{equalcontribution}}
\author[1,5]{Robert Frysch}
\author[7]{Vojtěch Kulvait}
\author[1]{Vladimir Semshchikov}
\author[10,5]{Bennet Hensen}
\author[10,5]{Frank Wacker}
\author[11,5]{Inga Brüsch}
\author[10,5]{Thomas Werncke}
\author[4,5,8,9]{Oliver Speck}
\author[2,3,9]{Andreas N{\"u}rnberger}
\author[1,5,9]{Georg Rose}

\cortext[equalcontribution]{H. Haselji{\'c} and S. Chatterjee have Equal Contribution}

\address[1]{Institute for Medical Engineering, Otto von Guericke University Magdeburg, Germany}
\address[2]{Faculty of Computer Science, Otto von Guericke University Magdeburg, Germany}
\address[3]{Data and Knowledge Engineering Group, Otto von Guericke University Magdeburg, Germany}
\address[4]{Biomedical Magnetic Resonance, Otto von Guericke University Magdeburg, Germany}
\address[5]{Research Campus STIMULATE, Otto von Guericke University Magdeburg, Germany}
\address[6]{Genomics Research Centre, Human Technopole, Milan, Italy}
\address[7]{Institute of Materials Physics, Helmholtz-Zentrum hereon, Geesthacht, Germany}
\address[8]{German Centre for Neurodegenerative Disease, Magdeburg, Germany}
\address[9]{Centre for Behavioural Brain Sciences, Magdeburg, Germany}
\address[10]{Institute of Diagnostic and Interventional Radiology, Hannover Medical School, Hannover, Germany}
\address[11]{Institute for Laboratory Animal Science, Hannover Medical School, Hannover, Germany}

\begin{abstract}
Model-based reconstruction employing the time separation technique (TST) was found to improve dynamic perfusion imaging of the liver using C-arm cone-beam computed tomography (CBCT). To apply TST using prior knowledge extracted from CT perfusion data, the liver should be accurately segmented from the CT scans. Reconstructions of primary and model-based CBCT data need to be segmented for proper visualisation and interpretation of perfusion maps. This research proposes Turbolift learning, which trains a modified version of the multi-scale Attention UNet on different liver segmentation tasks serially, following the order of the trainings CT, CBCT, CBCT TST - making the previous trainings act as pre-training stages for the subsequent ones - addressing the problem of limited number of datasets for training. For the final task of liver segmentation from CBCT TST, the proposed method achieved an overall Dice scores of 0.874±0.031 and 0.905±0.007 in 6-fold and 4-fold cross-validation experiments, respectively - securing statistically significant improvements over the model, which was trained only for that task. Experiments revealed that Turbolift not only improves the overall performance of the model but also makes it robust against artefacts originating from the embolisation materials and truncation artefacts. Additionally, in-depth analyses confirmed the order of the segmentation tasks. This paper shows the potential of segmenting the liver from CT, CBCT, and CBCT TST, learning from the available limited training data, which can possibly be used in the future for the visualisation and evaluation of the perfusion maps for the treatment evaluation of liver diseases.
\end{abstract}


\end{frontmatter}


\section{Introduction}
Computed Tomography (CT) perfusion or CTp imaging is a method that can be used for the diagnosis and treatment planning of liver tumours. C-arm cone-beam CT, referred to here in short as CBCT, on the other hand, can be advantageous during interventions as the acquisitions can be done without moving the patient due to the availability of CBCT as a part of the interventional suites~\citep{orth2008c}. It has been shown that CBCT perfusion maps of the brain would not be inferior to the CT perfusion maps~\citep{Niu2016-lx}, and when CT perfusion scans are acquired soon enough, it could the patient's life~\citep{powers2019guidelines}. C-arm CBCT perfusion (CBCTp) imaging of the liver could allow inspection and evaluation of the performed tumour embolisation right after it was performed without the hassle of transferring the patient. However, it is not yet part of standard procedures. Potentially it might also serve for diagnosing liver diseases. 
The experimental C-arm CBCTp scanning protocol of the liver consists of multiple bidirectional rotations with pauses in between~\citep{Datta2017}, which, combined with slow rotation, results in a very limited number of projections. A simplified approach would be to reconstruct every rotation separately, the straightforward approach, which 
can result in over or underestimation of perfusion parameters~\citep{Haseljic2021}. Recent publications have shown that model-based reconstruction and time separation technique (TST) could deal with poor temporal resolution~\citep{Montes2009-lp, Neukirchen2010, Manhart2013-yw, Bannasch2018,Kulvait2022,Haseljic2021, Haseljic2022} and provide highly accurate liver perfusion maps. 

\begin{figure}
    \centering
    \includegraphics[width=0.48\textwidth]{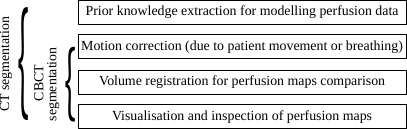}
    \caption{Downstream tasks for liver segmentation from CT, CBCT straightforward and CBCT model-based (TST) reconstructions.}
    \label{fig:usage}
\end{figure}

With TST, the time attenuation curve (TAC) of each voxel is modelled as a linear combination of orthonormal basis functions. The perfusion parameters (i.e. blood flow, blood volume, mean transit time, time to peak) are estimated based on the reconstructed coefficients. 
TST can incorporate prior knowledge about typical contrast agent dynamics by means of an orthogonal basis function set, which can be extracted from previous CT perfusion measurements from multiple subjects by applying singular value decomposition (SVD). To extract the basis function set from the CT reconstructed volumes that would be capable of modelling the temporal changes inside the liver, all except the liver tissue (i.e. different tissues, bones, and instruments surrounding the liver) have to be ignored. Manually segmenting the required volumes can be a possible solution. However, that has scalability issues (i.e. if new volumes are to be added to compute the basis function set - to improve the quality of the SVD, segmenting manually each of them would be challenging. Hence, accurate automatic segmentation of the liver from the CT volumes is of utmost importance. Moreover, to visualise and quantitatively evaluate the perfusion maps estimated based on the TST coefficients, the liver has to be segmented from the reconstructed TST coefficients and C-arm CBCT volumes. This warrants 3D-3D volumetric registration also with CT, which could also benefit from the liver segmentation~\citep{avants2009advanced,chatterjee2022micdir}. The potential usability of liver segmentation is summarised in Fig~\ref{fig:usage}. What makes the automatic liver segmentation a challenging task is the similar attenuation values of the liver and other surrounding organs, which consequently affects the estimation of liver contours~\citep{Lu2018}. This is especially problematic for CBCT volumes due to limited low-contrast resolution~\citep{Miracle2009}, truncation artefacts and more noisy data. Since the CT and CBCT reconstructed volumes are time-dependent, and each volume represents the organ in different time instances, the vessels' branch visibility will differ as the contrast agent flows through the organ. This is different from the CBCT model-based reconstruction as explained in Sec~\ref{sec:methods}. Tumours and lesions can, but do not necessarily, have different attenuation coefficients than the normal liver, so the segmentation results based on thresholds can be sensitive~\citep{Hendee2014}. For performing automatic segmentation, deep learning models are commonly used~\citep{kavur2021chaos}. However, one of the hurdles in training such a model is the typical requirement of large training datasets, and compiling such datasets for the aforementioned modalities is a challenge. This research tries to mitigate this problem with the help of the proposed "Turbolift" learning.

\subsection{Related Work}
Deep learning based techniques have been widely used for various image segmentation tasks, including abdominal segmentation~\citep{kavur2021chaos}. Since the inception of the UNet model~\citep{ronneberger2015u}, UNet and its flavours started dominating the research community. \citet{zeng20173d} introduced a deep supervision mechanism (i.e. calculating loss in different scales) to improve the gradient flow within the network, consequently improving the training process and the final output. \citet{oktay2018attention} improved the skip-connections by adding attention gates. \citet{abraham2019novel} combined both the earlier approaches into one network model while also adding multi-scale input. Even with the developments in terms of the network architectures, one inherent problem still remains - training of deep learning models typically requires large training datasets - finding such datasets for the types of data mentioned earlier is a challenging task. If the training set is significantly different from the test set, the model might not be able to perform well on the test set~\citep{wang2018deep,wilson2020survey}. Techniques, such as Data augmentation~\citep{perez2017effectiveness}, synthetic data generation~\citep{lateh2017handling,frid2018gan}, even domain adaptation~\citep{wang2018deep,wilson2020survey} can be employed to improve the model's performance on the test set while being trained on a small training dataset. But these techniques artificially modify the data, which might not be ideal or desirable when it comes to medical images. Transfer learning can be employed for adapting a pretrained model with fine-tuning~\citep{bengio2017deep} and fine-tuning with an anatomically similar dataset to the test set without artificially modifying the data, which has been seen to improve the performance of the model for the task of super-resolution while being trained on a dataset which is very different from the test set~\citep{sarasaen2021fine}. 

The importance of the liver segmentation for liver cancer treatment has been a subject of research \citep{Beichel2012-wq,Manjunath2022-ra}. Several studies have been published addressing the issue of liver segmentation in perfusion MRI~\citep{chen2008improved,dura2018method}, and contrast-enhanced CT~\citep{Draoua2014-ry,Christ2016-zh,Beichel2012-wq}. But liver segmentation in CBCT has not been explored enough. \citet{jafargholi2019multi} presented a semi-automatic method for liver segmentation from multi-modal perfusion data - PET/CT, SPECT/CT, and CBCT. However, they employ registration as a precursor to segmentation - making it not suitable for real-time applications, which interventions would demand.   

\subsection{Contributions}
This paper addresses the problem of liver segmentation from two different types of perfusion imaging modalities - CT and CBCT. Moreover, it performs segmentation from two different types of reconstructions of CBCT - straightforward and model-based (TST). To tackle the lack of large training datasets, this paper proposes the "Turbolift" learning technique - in which the model learns in different stages on CT, CBCT, and CBCT TST, where the earlier stages act as pre-training for the subsequent stages while also providing the segmentation output at each stage. This paper performs two different sets of cross-validation experiments - 4-fold and 6-fold and three different in-depth analyses to gain more insights regarding the proposed method. This paper also provides a comprehensive overview of CBCT reconstructions - straightforward and model-based time separation technique.

\section{Methodology}
\label{sec:methods}
This section presents the methodology used in this research. It starts by presenting the dataset - including a comprehensive overview of the CBCT model-based reconstruction method - time separation technique. Afterwards, the network architecture is introduced, and the working mechanism of "Turbolift" learning is presented. Then the information related to the implementation, including data augmentation techniques, is furnished. Finally, the evaluation metrics are presented.

\begin{figure*}
\centering
    \includegraphics[width=\textwidth]{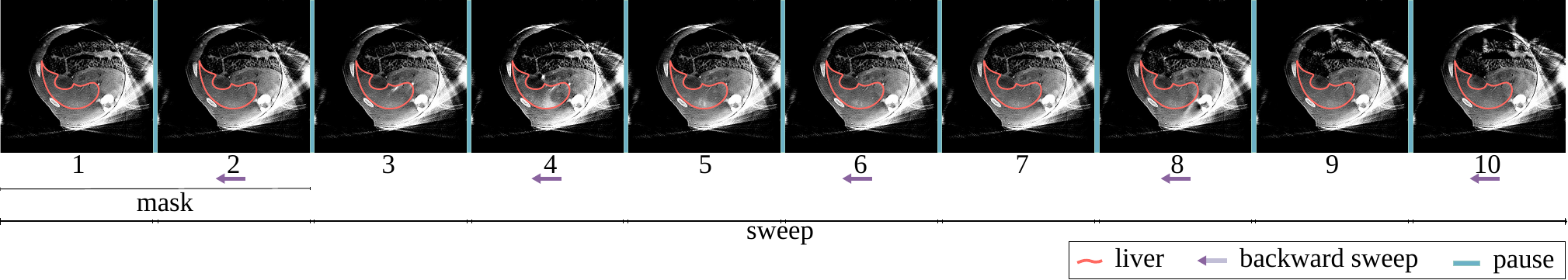}
    \caption{CBCT perfusion scan consists of ten rotations, five forward and five backward, with pauses between every two rotations. Volume is reconstructed for every rotation by assuming all projections to be acquired at the same time - straightforward reconstruction.}
    \label{fig:cbct}
\end{figure*}

\subsection{Dataset and Preprocessing}

CT and C-arm CBCT perfusion scans (CTp and CBCTp) of the liver of four different pigs \footnote{The study was conducted in accordance with the European Directive 2010/63/EU and with the German law for animal protection (TierSchG). All experiments were approved by the local animal ethic committee (Lower Saxony State Office for Consumer Protection and Food Safety, LAVES 18/2809).} were acquired after conducting embolisation (i.e. blockage) of the suitable subsegmental liver artery using tantalum-based embolisation material, Onyx\textsuperscript{\tiny\textregistered}, Medtronic, Tolochenaz, France in three subjects and Gelfoam in one. The latter one is used since it is not visible in CT.  
The embolisation results in decreased blood flow to liver tumour regions. 
The CTp scans were acquired using a Siemens SOMATOM Force CT and CBCT using a Siemens Artis pheno C-arm. All CTp and CBCTp scans were performed ten minutes apart, guaranteeing that no contrast agent of the previous scan was left. The CT perfusion scan was followed by the analytical reconstruction in syngo CT VA50A software with the slice thickness for all subjects \SI{1.5}{\milli\metre} except for one \SI{3.0}{\milli\metre} and the CBCT perfusion scans by the straightforward and model-based reconstruction with the slice thickness \SI{1}{\milli\metre}.

\subsubsection{C-arm Cone Beam CT Perfusion Scan}
\label{sec:perfscan}

The C-arm rotates around the patient, with the source on one side, which is directing the cone-beam shaped X-rays on the detector on another side of the C-arm. 
The C-arm CBCTp scan consists of ten sweeps - acquired with a total scan duration of 52 seconds, forward-backward pairs, with a pause between each two as shown in Fig.~\ref{fig:cbct}. Considering the slow C-arm rotation, this results in poor temporal resolution. Each sweep covered rotation of \SI{200}{\degree} and with the angular step of \SI{0.8}{\degree} acquired 248 projections. This is referred to as the longer C-arm CBCTp protocol in~\citet{Datta2017} where it was shown that the number of projections per sweep affected the results. Thus, the longer CBCTp scan protocol was also used in~\citet{Haseljic2021, Haseljic2022}. The first two sweeps are mask sweeps - meaning that there is yet no contrast agent present in liver vessels. The peak in arterial signal was present in the fourth sweep. During this, third, and fifth sweeps, the vessel signal was more pronounced than in the last sweeps during the contrast agent outflow.

\begin{figure}
    \centering
    \includegraphics[width=\columnwidth]{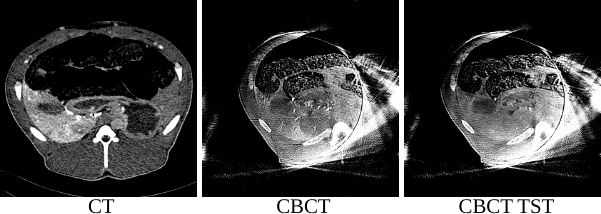}
    \caption{Reconstructed volumes for CT and CBCT and reconstructed first coefficient for CBCT TST. The CBCT TST reconstructed coefficient does not model the dynamic behaviour of the organ. Therefore, the different attenuation values of the liver tissue and the vessels cannot be observed as in CBCT volume given in the middle.}
    \label{fig:diff1b}
\end{figure}

\subsubsection{Straightforward Reconstruction}

The CBCT is reconstructed using the straightforward and model-based approach. In the straightforward approach, all the projections from one sweep are treated as if they were acquired at the same time $t$ and $\mathcal{R}$ standard reconstructions are computed as:
\begin{equation}
    \mathbf{A x_i} = \mathbf{p_i} \quad i \in \mathcal{R} 
\label{eq:stat}
\end{equation}
where $A$ is a system matrix, $x$ is a volume to reconstruct, $p$ is a projection vector and $R$ is a number of sweeps in CBCT scan. 

This means that for every scan, ten volumes per animal were obtained. The drawback of this reconstruction approach is that it doesn't correct the temporal undersampling of the data. All projections from one sweep cannot be considered to be acquired at the same time instance since this is not the case, considering the slow rotation and the angle coverage of the C-arm. To compensate for the missing data due to temporal undersampling, so far, only the TST has shown to give promising results with liver perfusion~\citep{Haseljic2022}. 


\subsubsection{Time Separation Technique}
\label{sec:tst}

As a C-arm makes its rotation around the patient at every time point $t$, only one projection for a certain angle is acquired. The reconstruction problem at the time $t$:

\begin{equation}
    \mathbf{A} \mathbf{x}(t) = \mathbf{p}(t) \quad t \in \mathcal{I}
\label{eq:init}
\end{equation}
where  $\mathcal{I}$ is a vector of time points at which the volume was scanned, is a time resolved problem since the projections from other angles at the same time $t$ are not acquired.

With TST, the CBCT data in the projection and reconstruction domain are modelled as a linear combination of the orthonormal temporal basis functions (Eq.~\ref{eq:basis}), 

\begin{equation}
    \mathcal{B} = \{\Psi_1, \ldots, \Psi_N\}
\label{eq:basis}
\end{equation}
reducing that way the number of reconstructions performed to the number of basis functions used $N$ ~\citep{Bannasch2018}. The time attenuation curve of each voxel $x_v$ takes the form

\begin{equation}
    x_v(t) = \sum_{i = 1}^{N} {w_{v,i}} \psi_i(t)
\label{eq:basis}
\end{equation}
and the projection of every pixel $p_n$
\begin{equation}
    p_n(t) = \sum_{j = 1}^{N} {\omega_{n,j}} \psi_j(t)
\label{eq:basis}
\end{equation} 

Now the reconstruction problem takes the form:

\begin{equation}
    \mathbf{A} \sum_{i = 1}^{N} \mathbf{w}_{i} \psi_i(t) = \sum_{j = 1}^{N} \boldsymbol{\omega}_{j} \psi_j(t) \quad t \in \mathcal{I}. 
\label{eq:tst}
\end{equation}

By multiplying both sides of the Eq.~\ref{eq:tst} with bases from $\mathcal{B}$ due to orthonormality, the TST simplifies model-based reconstruction problem to only $N$ standard reconstruction problems, given by:

\begin{equation}
    \mathbf{Aw}_{i} = \boldsymbol{\omega}_{i} \quad i \in N.
\label{eq:mbpr}
\end{equation}

\begin{figure}
    \centering
    \includegraphics[width=\columnwidth]{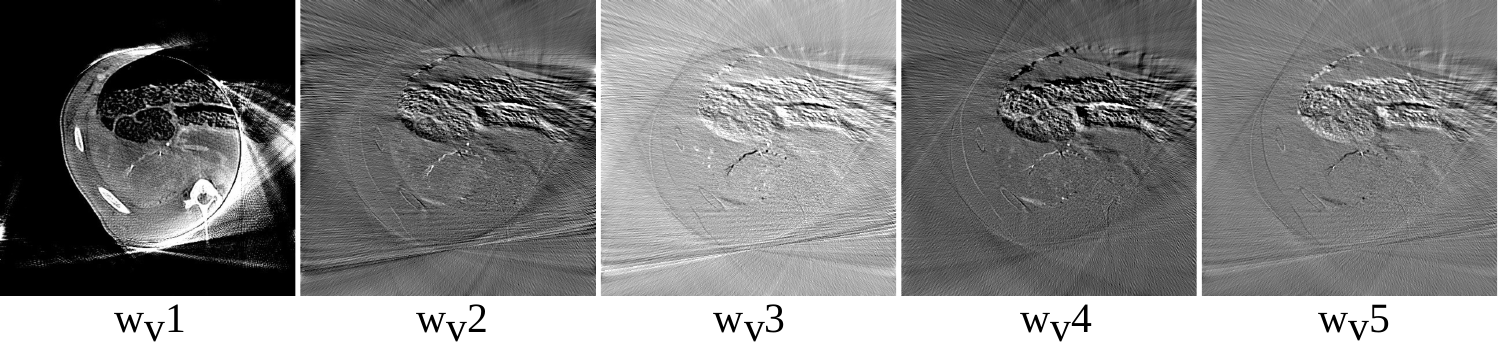}
    \caption{CBCT TST reconstructed coefficients. The basis function set was generated from CT reconstructions by applying singular value decomposition. Only the first five coefficients are selected to model liver perfusion.}
    \label{fig:coeff}
\end{figure}

\begin{figure}
    \centering
    \includegraphics[width=\columnwidth]{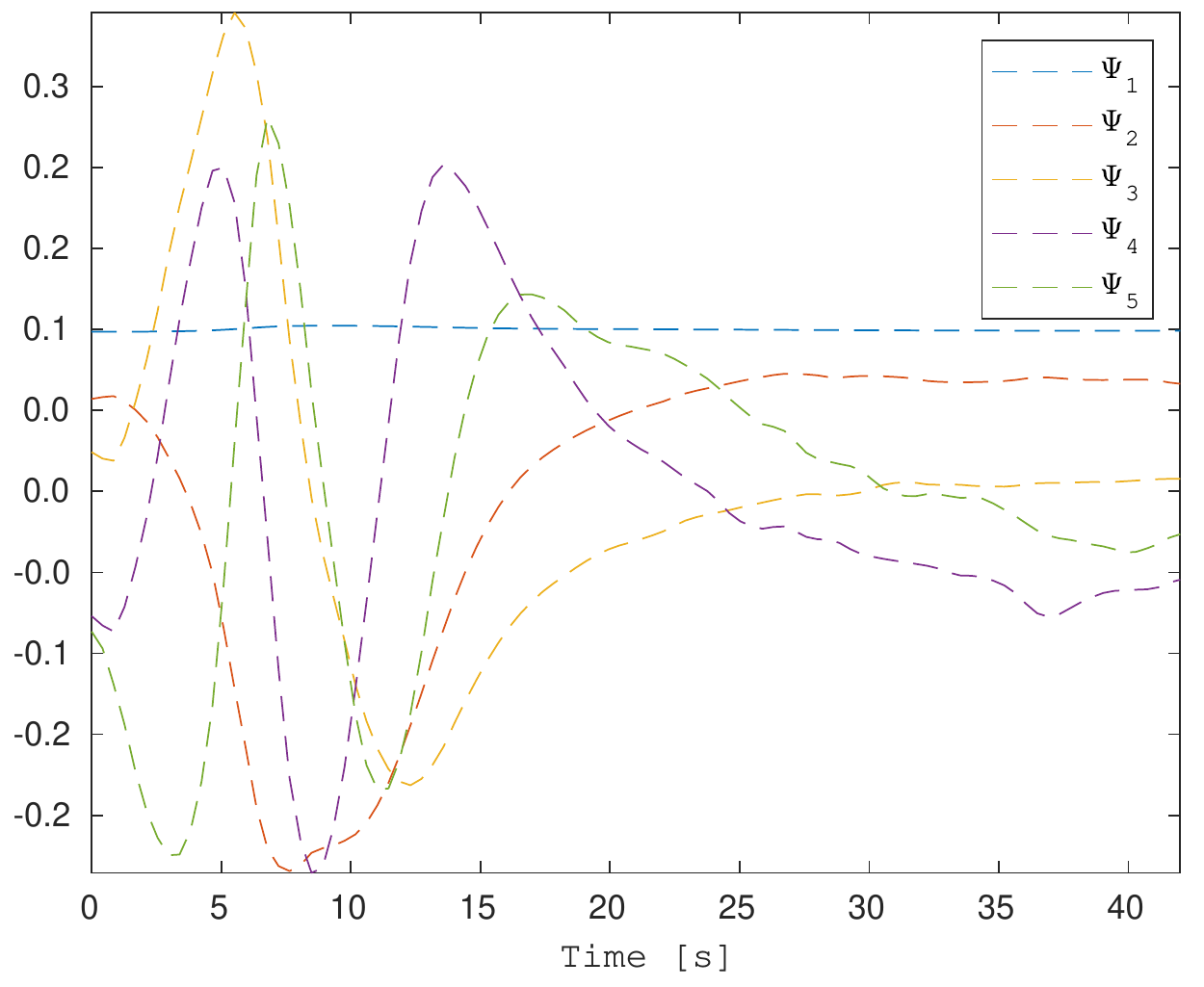}
    \caption{Orthonormal basis function set extracted from prior knowledge.}
    \label{fig:basissvd}
\end{figure}

The $\mathbf{w}$ coefficients are estimated by fitting basis functions to projection data and are used to estimate $\boldsymbol{\omega}$ coefficients~\citep{Kulvait2022}. In~\citep{Kulvait2022} only the first coefficient, which corresponds to a constant basis, is used for brain segmentation. When using the analytical basis functions, this is also the case for the liver segmentation. Here, the analytical basis function set is formed using five harmonic functions
\begin{equation}
\begin{split}
    \Psi_{0} = 1, \quad \Psi_1&=\sin(\frac{2\pi t}{T}), \quad \Psi_2=\cos(\frac{2 \pi t}{T}),\\ \Psi_3&=\sin(\frac{4 \pi t}{T}), \quad \Psi_4=\cos(\frac{4 \pi t}{T}).
\end{split}
\end{equation}
However, when using basis functions obtained by applying SVD on the CT volumes, the first coefficient is not a constant but it models the static behaviour as the most pronounced signal, see Fig.~\ref{fig:basissvd}. It is not easy to segment the liver from any other coefficient alone without composing it with the first one, as can be observed in Fig.~\ref{fig:coeff}.

\begin{figure}
    \centering
    \includegraphics[width=\columnwidth]{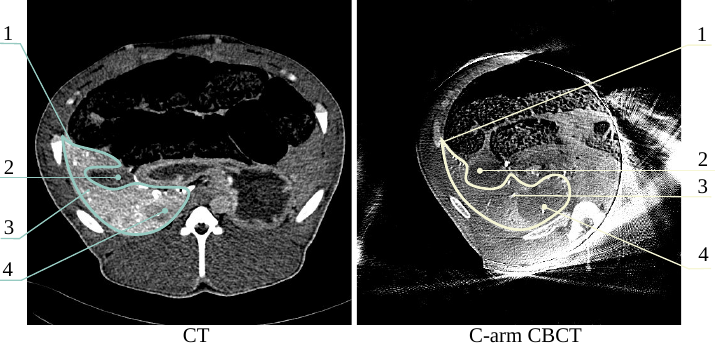}
    \caption{The liver annotation for the CT volume (left) and CBCT volume(right), 1: liver contours, 2: gallbladder, 3: vessels enhanced by contrast agent, 4: embolised/hypoperfused area.}
    \label{fig:diff}
\end{figure}

The liver was manually annotated in the volumes of the CT. CBCT, and CBCT TST datasets to be used as the ground-truth for training and evaluation. Depending on the contrast agent injection protocol, the vessels appear differently in the images making CT volumes more diverse in terms of vessel visibility during scan time than in CBCT images with much lesser number of volumes. With the TST using the SVD basis function set, usually the second bases represents the vessel signal, therefore there is a difference between how the vessels and differently perfused liver regions are reconstructed in volumes of the straightforward reconstruction and represented in the first coefficient, as shown in Fig~\ref{fig:diff1b}. All the mentioned were used for training and evaluation. In total, 52 CT, 37 CBCT, and 20 CBCT TST volumes were used in this research. Only in three sweeps of one CBCT scan the breathing was induced and these sweeps due to their exceptionality were excluded. To generate 20 CBCT TST volumes, different basis function sets were used to model data - including analytical and SVD sets.


\subsection{Network Architecture}
At the initial phase of this research, different deep learning architectures were experimented - UNet~\citep{ronneberger2015u}, UNet deep supervision~\citep{zeng20173d}, Attention UNet~\citep{oktay2018attention}, Multi-scale Attention UNet~\citep{abraham2019novel}, a modified version of the Multi-scale Attention UNet~\citep{chaosMEMoRIAL}, and UNet MSS~\citep{chatterjee2020ds6}. It is to be noted that the architectures, which were proposed for 3D, were modified to make them work in 2D. Among these, the best performing model in terms of the Dice score was the modified version of the Multi-scale Attention UNet~\citep{chaosMEMoRIAL} model, modified following the submission of the "OvGUMEMoRIAL" team at the CHAOS challenge~\citep{kavur2021chaos,chaosMEMoRIAL} - where the versatility of this architecture has been for multi-modal tasks. Given the aim of this paper is to segment the liver from multi-modal (CT and CBCT) perfusion data, while additionally dealing with two different types of CBCT reconstructions (straightforward and model-based TST), these findings are in-line with the previous findings~\citep{kavur2021chaos}. Moreover, this paper deals with one signal region of interest - the liver, which is typically located in a similar region in all the datasets - attention UNet or its flavours like the modified version of the Multi-scale Attention UNet~\citep{chaosMEMoRIAL} might be able to learn the focus area better due to the attention mechanism. Hence, this model was chosen for this research as the final model, and all the analyses were performed on this.

\begin{figure*}
    \centering
    \includegraphics[width=\textwidth]{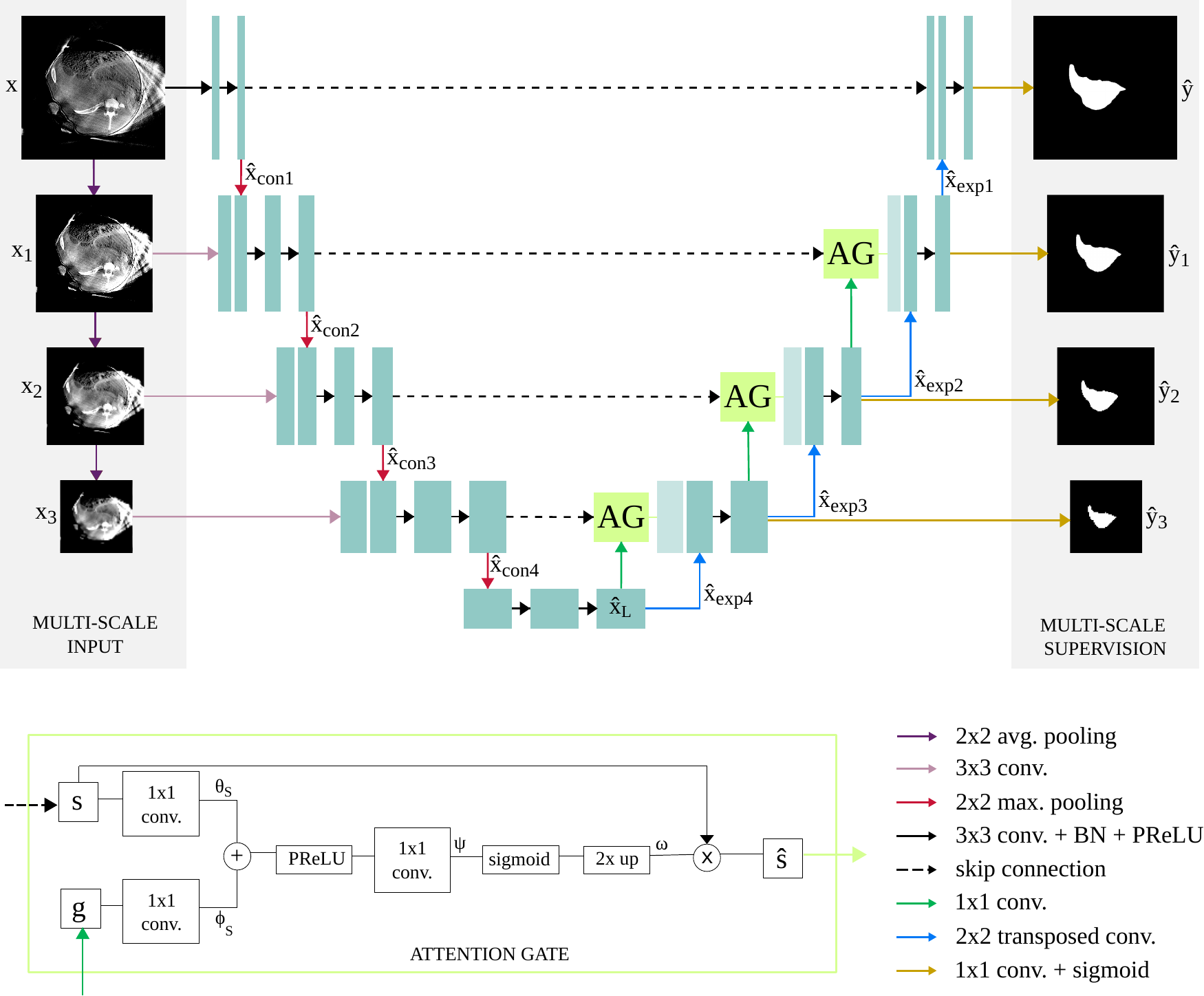}
    \caption{Multi-scale Attention UNet architecture: the input data is down-scaled using average pooling and supplied to the different stages of the contraction path, while the output of the model is obtained from the different stages of the expansion path super multi-scale supervision. The figure shows an example input slice and its corresponding output segmentation - including in different scales exactly how it was supplied to the network and how the network returned.}
    \label{fig:net}
\end{figure*}

The chosen network model~\citep{chaosMEMoRIAL} contains four downsampling stages and four upsampling stages, creating the contraction and expansion paths, respectively. The input $x$ first goes through two sets of convolutional layers with a kernel size of three, a stride of one, and a padding of one, batch normalisation, and PReLU activation function - referred to here as the "convolutional set" ($2 \times$ [Conv+BN+PReLU]) to obtain $\hat{x}_{con1}$. Then max pooling with a kernel of two is applied to reach the next stage to obtain $x_{con2}$, and again the convolutional set was applied, which generates $\hat{x}_{con2}$. In this manner, the input goes through all four stages. This model also receives input in multiple scales, i.e. input in different stages. For the same, the original input is downsampled in three different scales ($x_1$,$x_2$,$x_3$) with the help of average pooling with a kernel size of two. From the second to the fourth stage, this down-scaled input goes through a convolution layer with a kernel size of three, a stride of one, and a padding of one, and then gets concatenated with the output of the earlier stage ($\hat{x}_{con(n-1)}$) before going ahead with the convolutional sets. The output of the fourth stage $\hat{x}_{con4}$ also goes through the convolutional set in a similar manner, forming the latent space of the network $\hat{x}_{L}$. Afterwards, the expansion path of the network starts. The latent space is fed into a transposed convolution layer with a kernel size of two, a stride of two, and without any padding to reach the next stage of the expansion path, where the data goes through a convolutional set to obtain $\hat{x}_{exp1}$. Similarly, four stages of the expansion path are created. The output of each of the contraction path stages is concatenated with the output of each of the transposed convolutions as "skip connections" $s_n$, except for the final expansion path stage, after passing them through additive attention gets (AG). These gates first receive the input after sending the output of the earlier stage through a set of convolution layers with a kernel size of one, a stride of one, and without any padding to generate the gating signal $g$. The skip connection $s$ is passed through a convolution layer with kernel size and stride as two to generate $\theta_s$, and the gating signal is passed through a convolutional layer with kernel size and stride as one to obtain $\phi_g$. Then, PReLU activation is applied on $\theta_s + \phi_g$ before passing it through a fully-connected convolution layer with kernel size and stride as one which provides one signal channel output $\psi$. Afterwards, sigmoid activation is applied on $\psi$ before upsampling it with a scale factor of two in all dimensions using bilinear interpolation and is multiplied with $s$ to obtain $\omega$. Finally, a convolution with kernel size and stride of one, followed by batch normalisation, is applied to this interpolated data to generate the final output of the AG - $\hat{s}$. The output of each of the expansion path stages ($\hat{x}_{exp1}$,$\hat{x}_{exp2}$,$\hat{x}_{exp3}$, and $\hat{x}_{exp4}$) goes through one fully-connected convolution each (kernel size and stride are one), followed by sigmoid activations, to obtain the final output of the different stages, resulting in $\hat{y}_3$,$\hat{y}_2$,$\hat{y}_1$, and $\hat{y}$. The network architecture is shown in Fig.~\ref{fig:net}. This $\hat{y}$ is the final output of the model - the final segmentation, which is also used for all the evaluations. During training, the loss is calculated by comparing all these outputs against the different scaled version of the ground-truth - $y_3$,$y_2$,$y_1$, and $y$, where $y$ is the actual ground-truth and the different scaled versions are created by downsampling the $y$ with a factor of 2 in both dimensions using the nearest neighbour interpolation ($NN$, $NN_{-2}$ implies downsampling using $NN$ by a factor of 2) repeatedly - $y_1=NN_{-2}(y),y_2=NN_{-2}(y_1),y_3=NN_{-2}(y_2)$. This is known as the deep or multi-scale supervision, where the total loss ($\mathcal{L}_{mss}$) is calculated by summing up the loss at different scales, given by Eq.~\ref{eq:mss}. Each stage of the contraction path doubles the number of input feature maps, whereas the expansion path stages halve the number. At the initial stage, the output of the first stage is 64 feature maps. All the PReLU activations hold the same number of trainable parameters $\alpha$ as the number of output feature maps as the preceding convolution layer, resulting in different $\alpha$ for each of the feature maps in the network.

\begin{equation} 
\label{eq:mss}
\mathcal{L}_{mss} = \frac{1}{m+1}\left(\ell(\hat{y}, y)+\left(\sum\limits_{i=1}^m \ell(\hat{y}_i, y_i)\right)\right)
\end{equation}  
where $m$ is the number of additional scales excluding the original scale (in this research, $3$), $\ell$ is the loss function, $\hat{y}$ is the network's final prediction (at the original scale), $y$ is the ground-truth, $\hat{y}_i$ is the prediction at the given scale $i$, and $y_i$ is the down-scaled ground-truth for the scale $i$. 

\subsection{Turbolift Learning: Working Mechanism and Theory}
\label{sec:turbolift}
\begin{figure}
    \centering
    \includegraphics[width=\columnwidth]{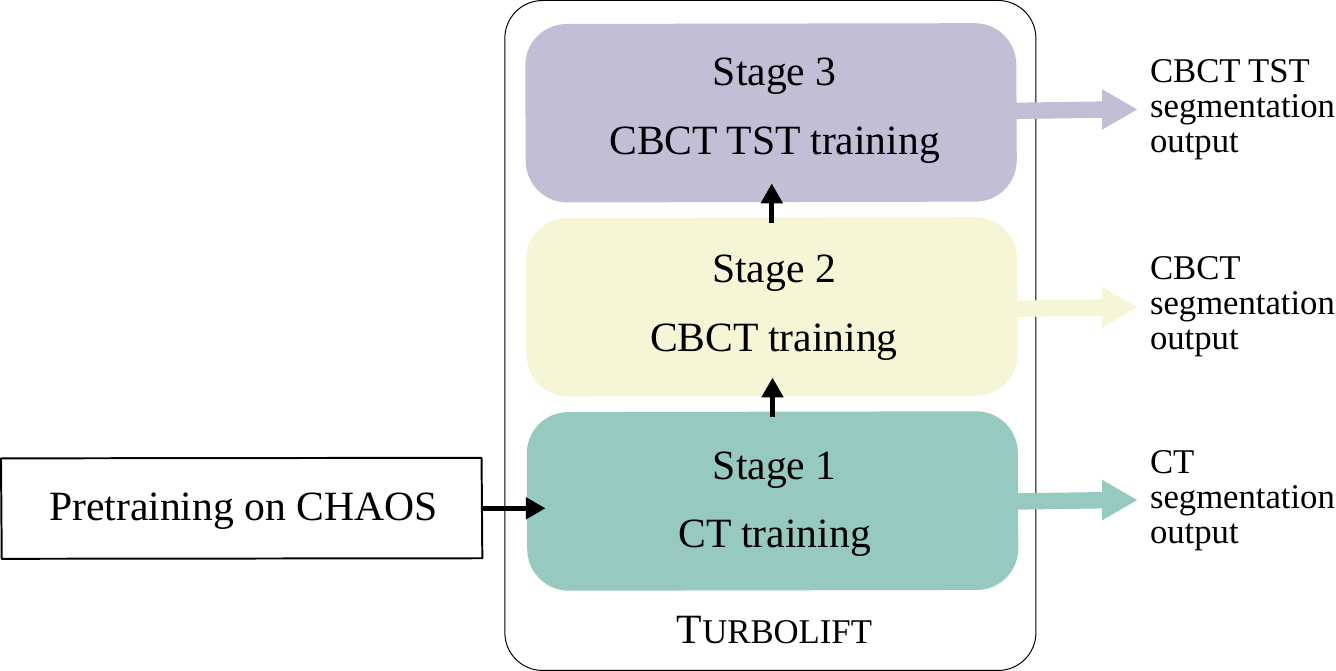}
    \caption{Turbolift learning: Training the model in different stages using different datasets to obtain the trained model weights and using them for segmenting that particular dataset (segmentation output during inference) at that stage, then using the earlier training as pre-training to train and infer on the next stage. Before the initial stage of CT, the model was pre-trained on a public dataset.}
    \label{fig:turbolift}
\end{figure}
Deep learning predominantly demands large training datasets similar to the final test set. Fine-tuning with an anatomically similar dataset can improve the performance of the model without artificially modifying the dataset~\citep{sarasaen2021fine}. Given the availability of only four animals and the data is of multiple modalities (CT and CBCT), including two different types of CBCT reconstructions (straightforward and model-based), this theory of transfer learning was employed at multi-stages. Pre-training the network model, even on a completely unrelated dataset, instead of initialising the network weights randomly has been seen to improve the performance of the network on the target task~\citep{bengio2017deep,chatterjee2022classification}. Hence, the model was pretrained on the CHAOS CT dataset of healthy human beings~\citep{kavur2021chaos} before starting the actual training stages. At the first stage of the main training, the CHAOS pretrained model was trained on the CT perfusion dataset of the animals. After the model was trained, the next stage of training was performed - on the CBCT dataset. Finally, at the third and final stage, the CBCT trained models were trained on the CBCT TST dataset. In this manner of step-wise training, the model learns on different datasets at different stages - enabling the model to achieve good performance on the final and the smallest dataset - CBCT TST. For every subsequent task, the previous training stages act as the pre-training for that task, enabling the model to obtain three stages of pre-training (CHAOS, Animal CT, CBCT) before learning on the CBCT TST dataset. This method of training in different stages is termed here as the "Turbolift", as at every stage, the model performs segmentation on a particular dataset, while acting as the pre-training for the subsequent stages. The main idea behind Turbolift learning is to start with the easiest and the largest dataset, then continue to learn on more challenging datasets, which are more complicated in terms of the clarity of the region of interest in the data and even the size of the dataset. In this research, the first stage of the main training was performed on the CT dataset - which not only is the largest available dataset, but also the delineation between the liver (region of interest) and the surrounding tissues is the clearest. Based on the complexity of the data in terms of clarity of visualisation of the region of interest and the size of the dataset, the next two stages of the training were chosen to be CBCT and CBCT TST.     

\subsection{Implementation, Training, and Inference}
The model was trained on 2D slices from the training datasets in different stages, as explained earlier, with a batch size of eight while accumulating gradients of eight batches before backpropagating - creating an effective batch size of 64. The loss $\ell$ between the model's predictions and the ground-truth segmentation masks was calculated with the help of the focal Tversky loss~\citep{abraham2019novel}, defined by:

\begin{equation} 
\label{eq:ftl}
\mathcal{FTL} = (1-\mathcal{TI})^\frac{1}{\gamma}
\end{equation}

where $\gamma$ is to control the network's focus on the misclassifications that was set to $\frac{4}{3}$ here, and $\mathcal{TI}$ is the Tversky index, which is calculated using:

\begin{equation} 
\label{eq:ti}
\mathcal{TI} = \frac{\sum\limits_{i=1}^N p_{il}g_{il} + \epsilon }{\sum\limits_{i=1}^N p_{il}g_{il} + \alpha \sum\limits_{i=1}^N p_{i\bar{l}}g_{il} + \beta \sum\limits_{i=1}^N p_{il}g_{i\bar{l}} + \epsilon }
\end{equation}

where $p_{il}$ is the probability of the pixel $i$ belongs to the liver class, $p_{i\bar{l}}$ is the probability of the pixel not belonging to the liver class, same holds for the ground-truth $g_{il}$ and $g_{i\bar{l}}$, $\alpha$ and $\beta$ are hyperparameters which can be adjusted to tune the network's sensitivity - especially when there is a class imbalance in the dataset - set in this research to $0.7$ and $0.3$ respectively, and finally $\epsilon$ is to provide numerical stability - to avoid division by zero.  

\begin{figure}
    \centering
    \includegraphics[width=0.48\textwidth]{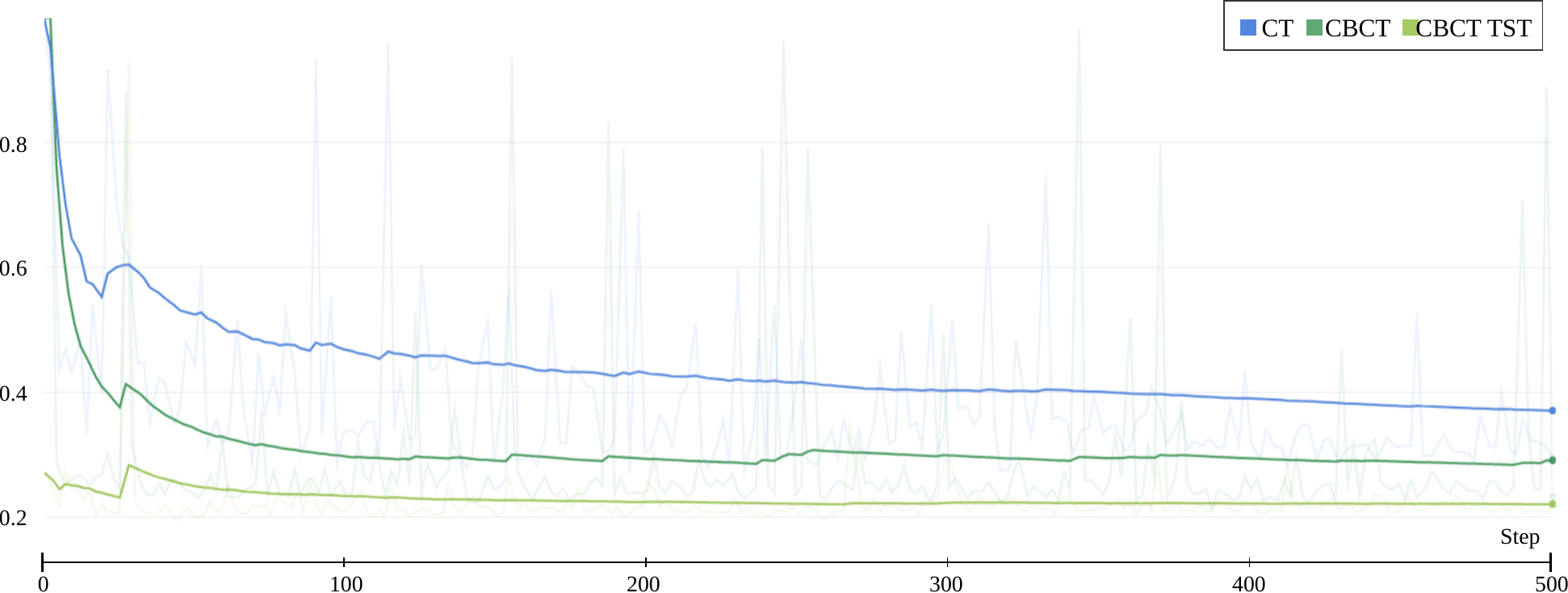}
    \caption{Validation loss graphs for the three stages of Turbolift learning - for the first fold of the 6-fold experiments}
    \label{fig:valloss}
\end{figure}

The loss $\ell$ was optimised using the Adam optimiser with a learning rate of 0.001 for 500 epochs. The training of the different stages was performed till convergence - which was much earlier than 500 in every case. To verify whether or not having a fixed number of epoch for all stages is a current strategy, validation losses for one of the experiments is shown in Fig.~\ref{fig:valloss}, and it confirms 500 epochs to be a sufficient amount of training for all the stages. At every stage of Turbolift, the trained models were used for inference on that particular dataset and then were treated as pre-training to perform training on the next. 6-fold cross-validation taking two animals each for training and validation, and 4-fold cross-validation taking three animals for training and one for validation, were performed at every stage of the training. Slices of one animal can be either used only for training or only for testing - otherwise the network could learn liver shape of specific animal and predict based on this in testing phase. The approach was implemented using PyTorch~\citep{NEURIPS2019_9015}, with the help of PyTorch~Lightning~\citep{falcon2019pytorch}; and was trained with mixed precision~\citep{micikevicius2018mixed} using Nvidia RTX 2080Ti GPUs. The code of this project is available on GitHub\footnote{Turbolift Code:~\url{https://github.com/soumickmj/Turbolift}}.

\subsubsection{Data Augmentation}
A commonly used technique for dealing with the problem of small training datasets is data augmentation, which randomly modifies the training data during every epoch - to provide a large variety of examples to the model over epochs. A set of four different data augmentation techniques, which do not modify the data in an unrealistic fashion, were chosen for this research, comprising random horizontal flips with a probability of $80\%$, random vertical flips also with a probability of $80\%$, random rotation with an angle between $-45^{\circ}$ and $45^{\circ}$ clockwise, and random translation of the pixels in both dimensions (height and width) between $-32$. This set of data augmentation techniques was applied during training with a probability of $75\%$.

\subsubsection{Postprocessing}
To get rid of unnecessary noise from the predictions, a set of postprocessing techniques were applied. After obtaining the predictions from the model, the different connected regions in every 2D prediction were labelled with unique labels. Then the area (in terms of the number of pixels) for each of those uniquely-labelled regions was calculated. Given the domain knowledge that there can only be one liver, the region with the largest area was chosen, and the rest were considered noise - hence, removed. 

\subsection{Evaluation Criteria}
The segmentation performance of the models was compared against the manually segmented ground-truth using two different quantitative metrics. The first metric which was employed was the Sørensen–Dice coefficient or Dice similarity coefficient (DSC) - commonly known as the Dice score, given by the following equation for Boolean data :

\begin{equation}
    DSC={\frac {2TP}{2TP+FP+FN}}
\end{equation}

where $TP$ signifies the number of true positives, $TN$ is the number of true negatives, $FP$ is the number of false positives, and $FN$ is the number of false negatives. The second metric used in this research was the Jaccard index, also known as the intersection over union (IoU) - which provides less importance to the true positives (counts them once in numerator and denominator), which is given by the following equation for binary classification tasks :

\begin{equation}
    {\text{IoU}}={\frac {TP}{TP+FP+FN}}
\end{equation}

Furthermore, as an image segmentation task can also be viewed as a pixel-level classification task, three classification metrics were used to evaluate the models' performance in terms of pixel-level classification. The first of these three metrics were precision ($\frac {TP}{TP+FP}$), which gives information regarding how many of the positively classified pixels were actually relevant. The next metric was sensitivity ($\frac {TP}{TP+FN}$), also known as hit rate or recall or true positive rate, which is a common metric used in classification to judge how good a classifier is at detecting the true positives. Finally, the last metric used was specificity ($\frac {TN}{FP+TN}$), also known as true negative rate, which provides information regarding the model's capability to avoid false negatives. This set of evaluation metrics was chosen as each one of them provides a different view of the results - from different perspectives.

The significance of the differences in the resultant metric values between the different methods was measured by the Mann-Whitney U-test~\citep{mann1947test,di1999combinatorics}.

\section{Results}

\subsection{6-Fold Cross-validation}

\bgroup
\def\arraystretch{1.5}
\begin{table*}[]
\centering
\caption{Quantitative comparison of the segmentation performances with and without Turbolift learning (with postprocessing) for 6-fold cross-validation using the median±variance Dice, intersection over union (IoU), precision, sensitivity, and specificity}
\label{tab:metrics_main_6fold}
\begin{tabular}{ccccccc}
Dataset                                                             & Turbolift & Dice        & IoU         & Precision   & Sensitivity & Specificity \\ \hline \hline
CT                                                                  & Yes/No    & 0.864±0.007 & 0.761±0.012 & 0.851±0.009 & 0.907±0.012 & 0.989±0.0   \\ \hline
\multirow{2}{*}{CBCT}                                               & No        & 0.853±0.010 & 0.744±0.014 & 0.836±0.013 & 0.900±0.014 & 0.989±0.0   \\
                                                                    & Yes       & \textbf{0.867±0.012} & \textbf{0.766±0.016} & \textbf{0.851±0.012} & \textbf{0.911±0.019} & \textbf{0.991±0.0}   \\ \hline
\multirow{2}{*}{\begin{tabular}[c]{@{}c@{}}CBCT\\ TST\end{tabular}} & No        & 0.858±0.031 & 0.752±0.030 & 0.845±0.028 & 0.905±0.039 & 0.991±0.0   \\
                                                                    & Yes       & \textbf{0.874±0.031} & \textbf{0.776±0.030} & \textbf{0.859±0.022} & \textbf{0.910±0.039} & \textbf{0.992±0.0}   \\ \hline
\end{tabular}
\end{table*}

\bgroup
\def\arraystretch{1.5}
\begin{table*}[]
\centering
\caption{Quantitative comparison of the segmentation performances with and without postprocessing (with Turbolift learning) for 6-fold cross-validation using the median±variance Dice, intersection over union (IoU), precision, sensitivity, and specificity}
\label{tab:metrics_postprocess_6fold}
\begin{tabular}{ccccccc}
Dataset & Postprocessed & Dice & IoU & Precision & Sensitivity & Specificity                  \\ \hline \hline
\multirow{2}{*}{CT} & No  & \multicolumn{1}{l}{0.841±0.006} & \multicolumn{1}{l}{0.725±0.009} & \multicolumn{1}{l}{0.801±0.008} & \multicolumn{1}{l}{\textbf{0.908±0.010}} & \multicolumn{1}{l}{0.985±0.0} \\
                    & Yes & \textbf{0.864±0.007}            & \textbf{0.761±0.012}            & \textbf{0.851±0.009}            & 0.907±0.012                              & \textbf{0.989±0.0} \\ 
\hline
\multirow{2}{*}{CBCT}   & No   & 0.854±0.012          & 0.745±0.016          & 0.817±0.013          & \textbf{0.912±0.018} & 0.988±0.0  \\
                        & Yes  & \textbf{0.867±0.012} & \textbf{0.766±0.016} & \textbf{0.851±0.012} & 0.911±0.019          & \textbf{0.991±0.0} \\ 
\hline
\multirow{2}{*}{\begin{tabular}[c]{@{}c@{}}CBCT\\ TST\end{tabular}} & No   & 0.862±0.030          & 0.757±0.029          & 0.840±0.023           & \textbf{0.910±0.037} & 0.990±0.0 \\
                                                                    & Yes  & \textbf{0.874±0.031} & \textbf{0.776±0.030} & \textbf{0.859±0.022}  & 0.910±0.039 & \textbf{0.992±0.0} \\ 
\hline
\end{tabular}
\end{table*}

\begin{figure}
    \centering
    \includegraphics[width=0.49\textwidth]{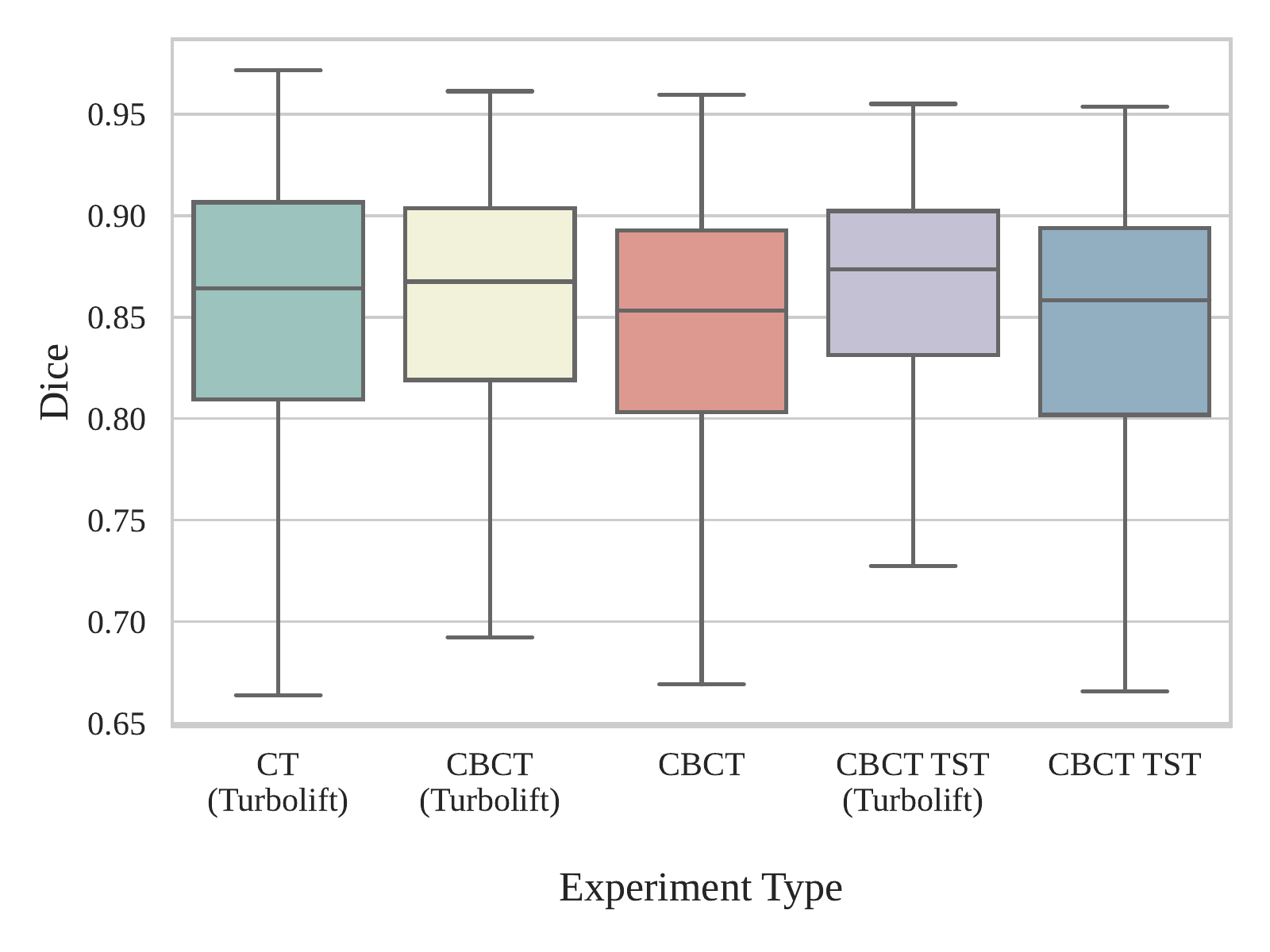}
    \caption{Dice scores for 6-fold cross-validation - comparing the Turbolift results against the results without Turbolift (i.e. models trained directly on that particular dataset)}
    \label{fig:Dice_results_6fold}
\end{figure}

Table~\ref{tab:metrics_main_6fold} presents the resultant metric values (in terms of median and variance) after comparing the final segmentation results (obtained after postprocessing the models' predictions) against the manually annotated ground-truth for segmenting liver from CT, CBCT, and CBCT TST datasets. To evaluate the contribution of Turbolift learning, the CHAOS-pretrained model was also trained on the CBCT and CBCT TST separately (without Turbolift learning). The segmentations generated after applying identical postprocessing techniques (as the Turbolift learning models) on the non-Turbolift models were then compared against the ground-truth - the resultant metric values are also presented in Table~\ref{tab:metrics_main_6fold} and were compared against the ones obtained by Turbolift learning. For CBCT, Turbolift achieved $1.64\%$, $2.96\%$, and $1.79\%$ improvements over the non-Turbolift model, in terms of Dice, IoU, and precision, respectively. Whereas for CBCT TST, Turbolift improved the Dice scores, IoU, and precision by $1.87\%$, $3.19\%$, and $1.66\%$, respectively. The statistical tests revealed that the improvements on all the five metrics achieved by Turbolift learning on both CBCT and CBCT TST datasets were statistically significant (p-values were always less than $0.05$). The resultant Dice scores are also visualised with the help of box plots in Fig.~\ref{fig:Dice_results_6fold}. A qualitative comparison of the results are presented in Figures~\ref{fig:results}~and~\ref{fig:results_GB} (second row in each).



Further comparisons were performed to evaluate the contribution of the postprocessing step in the final results. For the same, the Turbolift results are furnished in Table~\ref{tab:metrics_postprocess_6fold} where the scores obtained by the raw output of the models (without and postprocessing) were compared with the scores obtained after postprocessing them. Four out of five metrics (except sensitivity) show statistically significant improvements for all three types of data after applying postprocessing. However, in terms of sensitivity, a minor decrease can be observed for CT and CBCT, while the median sensitivity remains the same for CBCT TST while increasing the variance - all these differences were statistically insignificant.

\subsection{4-Fold Cross-validation}
Similar comparisons were also performed for the 4-fold cross-validation experiment. Table~\ref{tab:metrics_main_4fold} presents the resultant metric values (in terms of median and variance) after comparing the final segmentation results (obtained after postprocessing the models' predictions) against the manually annotated ground-truth for segmenting liver from CT, CBCT, and CBCT TST datasets, for Turbolift learning and for the models trained without Turbolift learning (CHAOS-pretrained model trained on the CBCT and CBCT TST separately). Postprocessing steps were identical for both types of trainings - with and without Turbolift learning. For CBCT, Turbolift achieved $2.99\%$, $5.47\%$, and $2.87\%$ improvements over the non-Turbolift model, in terms of Dice, IoU, and precision, respectively. Whereas for CBCT TST, Turbolift improved the Dice scores, IoU, and precision by $2.61\%$, $4.56\%$, and $0.80\%$, respectively. The statistical tests revealed that the improvements on all the five metrics achieved by Turbolift learning on CBCT and four out of five metrics (except specificity) on CBCT TST datasets were statistically significant. The resultant Dice scores are also visualised with the help of box plots in Fig.~\ref{fig:Dice_results_4fold}. A qualitative comparison of the results are presented in Figures~\ref{fig:results}~and~\ref{fig:results_GB} (third row in each). Furthermore, Table~\ref{tab:metrics_postprocess_4fold} presents the results of Turbolift learning, with and without postprocessing. Same as the 6-fold experiment, four out of five metrics (except sensitivity) show statistically significant improvements for all three types of data after applying postprocessing.



\bgroup
\def\arraystretch{1.5}
\begin{table*}[]
\centering
\caption{Quantitative comparison of the segmentation performances with and without Turbolift learning (with postprocessing) for 4-fold cross-validation using the median±variance Dice, intersection over union (IoU), precision, sensitivity, and specificity}
\label{tab:metrics_main_4fold}
\begin{tabular}{ccccccc}
Dataset                                                             & Turbolift & Dice        & IoU         & Precision   & Sensitivity & Specificity \\ \hline \hline
CT                                                                  & Yes/No    & 0.888±0.006 & 0.799±0.01 & 0.861±0.009 & 0.937±0.007 & 0.937±0.007   \\ \hline
\multirow{2}{*}{CBCT}                                               & No        & 0.869±0.009 & 0.768±0.013 & 0.837±0.012 & 0.923±0.01 & 0.989±0.0   \\
                                                                    & Yes       & \textbf{0.895±0.006} & \textbf{0.81±0.008} &  \textbf{0.861±0.007} & \textbf{0.939±0.008} & \textbf{0.991±0.0}   \\ \hline
\multirow{2}{*}{\begin{tabular}[c]{@{}c@{}}CBCT\\ TST\end{tabular}} & No        & 0.882±0.012 & 0.79±0.015 & 0.875±0.009 & 0.913±0.019 & 0.993±0.0   \\
                                                                    & Yes       & \textbf{0.905±0.007} & \textbf{0.826±0.01} & \textbf{0.882±0.005} & \textbf{0.921±0.011} & 0.993±0.0   \\ \hline
\end{tabular}
\end{table*}

\bgroup
\def\arraystretch{1.5}
\begin{table*}[]
\centering
\caption{Quantitative comparison of the segmentation performances with and without postprocessing (with Turbolift learning) for 4-fold cross-validation using the median±variance Dice, intersection over union (IoU), precision, sensitivity, and specificity}
\label{tab:metrics_postprocess_4fold}
\begin{tabular}{ccccccc}
Dataset & Postprocessed & Dice & IoU & Precision & Sensitivity & Specificity                  \\ \hline \hline
\multirow{2}{*}{CT} & No  & \multicolumn{1}{l}{0.865±0.005} & \multicolumn{1}{l}{0.763±0.01} & \multicolumn{1}{l}{0.826±0.009} & \multicolumn{1}{l}{\textbf{0.938±0.006}} & \multicolumn{1}{l}{\textbf{0.987±0.0}} \\
                    & Yes & \textbf{0.888±0.006}            & \textbf{0.799±0.01}             & \textbf{0.861±0.009}                     & 0.937±0.007                     & 0.937±0.007                \\ 

\hline
\multirow{2}{*}{CBCT}                                               & No             & 0.885±0.006          & 0.794±0.008         &  0.843±0.007          & \textbf{0.939±0.007}  & 0.990±0.0    \\
                                                                    & Yes            & \textbf{0.895±0.006} & \textbf{0.81±0.008} &  \textbf{0.861±0.007} & 0.939±0.008           &  \textbf{0.991±0.0}           \\ 
\hline
\multirow{2}{*}{\begin{tabular}[c]{@{}c@{}}CBCT\\ TST\end{tabular}} & No             & 0.9±0.006            & 0.819±0.01          & 0.871±0.006          & \textbf{0.922±0.01}  & 0.990±0.0               \\
                                                                    & Yes            & \textbf{0.905±0.007} & \textbf{0.826±0.01} & \textbf{0.882±0.005} & 0.921±0.011          & \textbf{0.993±0.0}      \\ 
\hline
\end{tabular}
\end{table*}

\begin{figure}
    \centering
    \includegraphics[width=0.49\textwidth]{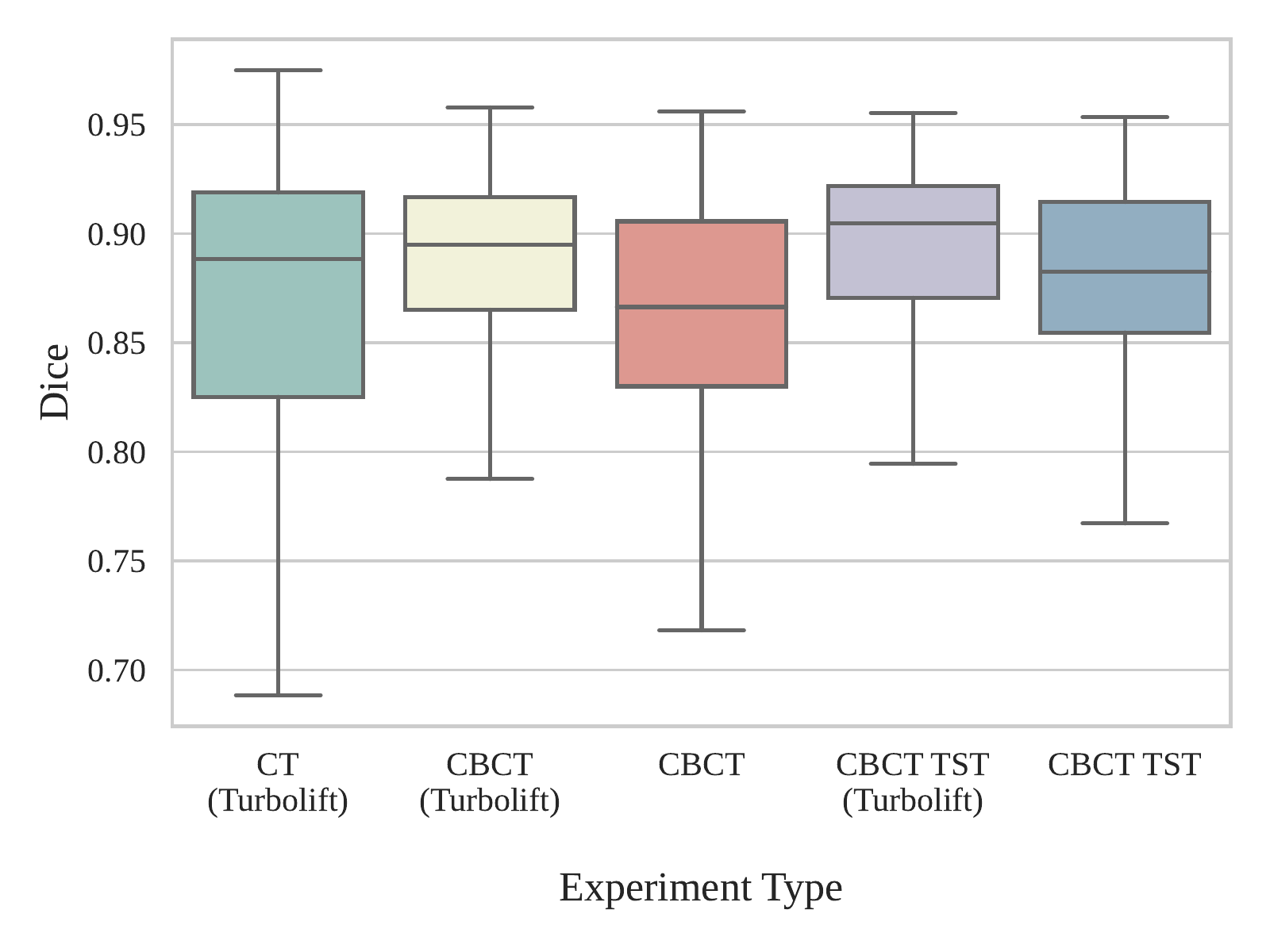}
    \caption{Dice scores for 4-fold cross-validation - comparing the Turbolift results against the results without Turbolift (i.e. models trained directly on that particular dataset)}
    \label{fig:Dice_results_4fold}
\end{figure}

\begin{figure*}
    \centering    \includegraphics[width=.9\textwidth]{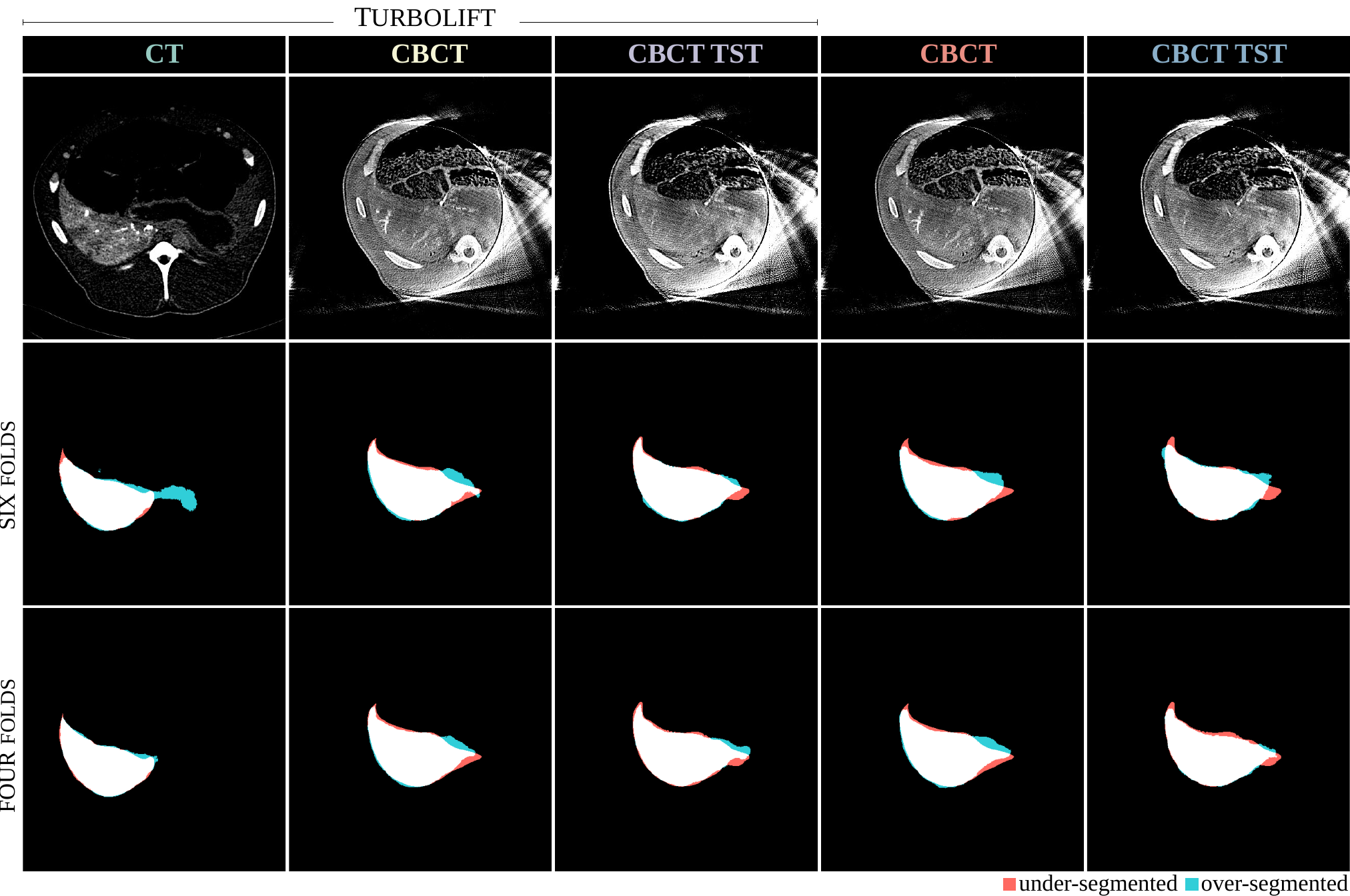}
    \caption{Qualitative comparisons of segmentations results with and without Turbolift (i.e. models trained directly on that particular dataset). The attenuation values of surrounding organs and tissue are very similar to attenuation value of liver, especially in the embolised area of the liver through which there is no contrast agent flow.}
    \label{fig:results}
\end{figure*}

\begin{figure*}
    \centering
    \includegraphics[width=.90\textwidth]{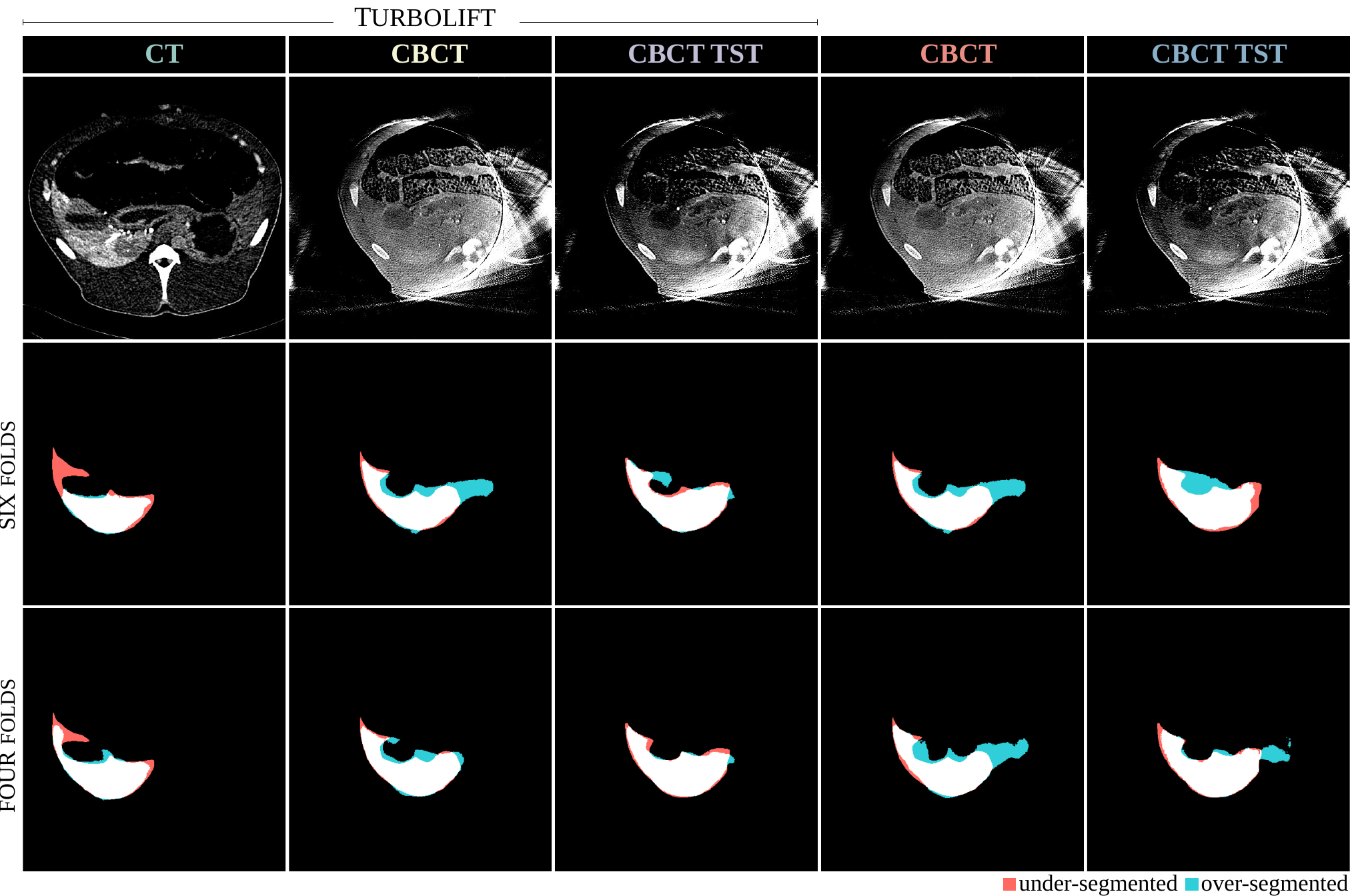}
    \caption{Qualitative comparisons of segmentations results with and without Turbolift (i.e. models trained directly on that particular dataset) for slices with gallbladder. The gallbladder is positioned right under the liver, making these slices very similar to other, and the attenuation values are, due to absence of contrast agent, similar to those of embolised  area in liver.}
    \label{fig:results_GB}
\end{figure*}

\subsubsection{In-depth analyses}
\label{sec:indepth}
Quantitative evaluations presented in Tables~\ref{tab:metrics_main_6fold}~and~\ref{tab:metrics_main_4fold} have shown that the 4-fold cross-validation resulted in higher scores, except for specificity on the CT dataset - where 6-fold resulted in a higher score and on the CBCT TST dataset - where both experiments resulted in the same values. Moreover, in most cases, the improvements observed on CBCT and CBCT TST datasets with Turbolift were more in the 4-fold experiments than with 6-fold. Hence, 4-fold cross-validation was chosen to obtain further insights regarding the method and the results on three different in-depth analysis scenarios - flipped Turbolift, reversed Turbolift, Turbolift without CHAOS pre-training, and testing the Turbolift models with artefacts. However, it is worth mentioning that given the difference in the size of the training and validation sets and also the number of folds, these experiments are not fully comparable.   

\paragraph{Flipped Turbolift}
The order of the Turbolift trainings was chosen based on the size of the dataset and other factors, as discussed in Sec.~\ref{sec:turbolift}. To confirm the hypothesis behind choosing this order, especially the order of training CBCT and CBCT TST, this experiment was performed - termed here as the Flipped Turbolift (fTL). For both TL and fTL, the model was initially pretrained on the CHAOS dataset and then trained on the CT dataset as the first stage of Turbolift (see Fig.~\ref{fig:turbolift}). Then for fTL, CBCT TST training was performed as the second stage, and finally, CBCT training was performed as the third stage. For TL, the order discussed in Fig.~\ref{fig:turbolift} was followed. The resultant metrics are presented in Table~\ref{tab:metrics_ftl_4fold}, while the Dice scores are visualised with the help of box plots in Fig.~\ref{fig:Dice_results_flpTL_4fold}. It can be seen that the TL resulted in higher scores than fTL in eight out of ten cases, and the statistical tests indicated them to be statistically significant. For CBCT, both TL and fTL resulted in the same median specificity. In the case of CBCT TST, fTL resulted in better sensitivity than TL - but the difference was statistically insignificant. In conclusion, these results support the chosen order for the Turbolift learning. 

\bgroup
\def\arraystretch{1.5}
\begin{table*}[]
\centering
\caption{Quantitative comparison of the segmentation performances with Turbolift and with flipped Turbolift (with postprocessing) for 4-fold cross-validation using the median±variance Dice, intersection over union (IoU), precision, sensitivity, and specificity}
\label{tab:metrics_ftl_4fold}
\begin{tabular}{ccccccc}
Dataset                                                             & Type & Dice        & IoU         & Precision   & Sensitivity & Specificity \\ \hline \hline
CT                                                                  & /    & 0.888±0.006 & 0.799±0.01 & 0.861±0.009 & 0.937±0.007 & 0.937±0.007   \\ \hline
\multirow{2}{*}{CBCT}                                               & fTL      & 0.885±0.004 & 0.794±0.008 & 0.855±0.007 & 0.935±0.007 & 0.991±0.0   \\
                                                                    & TL       & \textbf{0.895±0.006} & \textbf{0.81±0.008} &  \textbf{0.861±0.007} & \textbf{0.939±0.008} & 0.991±0.0   \\ \hline
\multirow{2}{*}{\begin{tabular}[c]{@{}c@{}}CBCT\\ TST\end{tabular}} & fTL        & 0.892±0.006 & 0.805±0.009 & 0.867±0.006 & \textbf{0.929±0.01} & 0.992±0.0   \\
                                                                    & TL       & \textbf{0.905±0.007} & \textbf{0.826±0.01} & \textbf{0.882±0.005} & 0.921±0.011 & \textbf{0.993±0.0}   \\ \hline
\end{tabular}
\end{table*}

\begin{figure}
    \centering
    \includegraphics[width=0.49\textwidth]{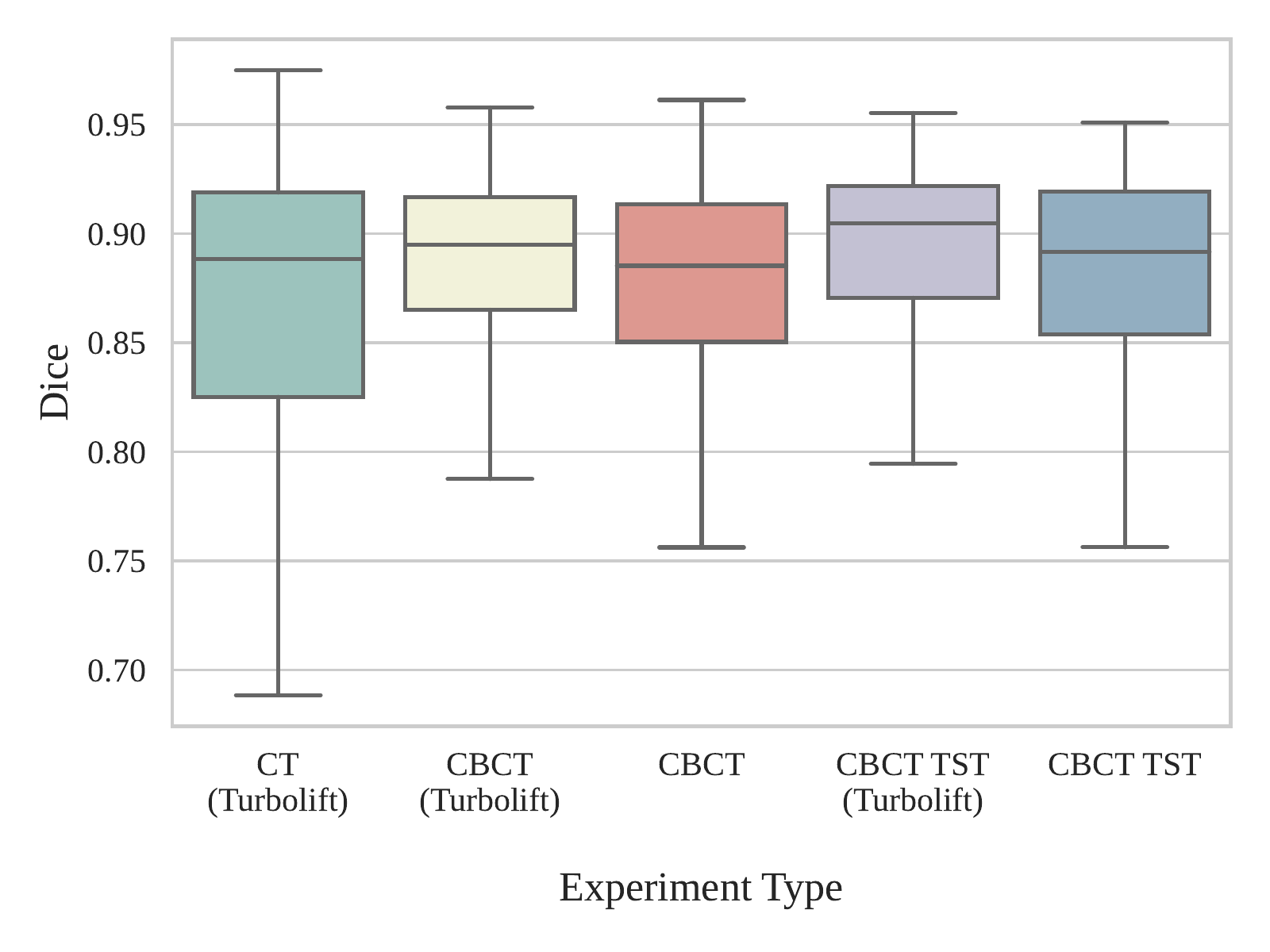}
    \caption{Dice scores for 4-fold cross-validation - comparing the Turbolift results against the results of flipped Turbolift}
    \label{fig:Dice_results_flpTL_4fold}
\end{figure}

\paragraph{Reversed Turbolift}
The chosen order was further evaluated by extending the flipped Turbolift experiment - by flipping the Turbolift order completely - starting with CBCT TST, followed by CBCT, and finally CT. This experiment is termed here as the Reversed Turbolift (rTL). In this experiment, the model was first pretrained on the CHAOS dataset before training the first stage of rTL (CBCT TST) and then continued on with the rest of the stages similar to the Turbolift. The resultant metrics are presented in Table~\ref{tab:metrics_rtl_4fold}, while the Dice scores are visualised with the help of box plots in Fig.~\ref{fig:Dice_results_rvrsTL_4fold}. These results further justify the initially selected order of the Turbolift, as it can be seen that all except precision and specificity on the CT dataset resulted higher scores with Tubolift than rTL. All, except three (Dice and IoU for CT, specificity for TST) of the differences in terms of the metric values in Table~\ref{tab:metrics_rtl_4fold} are statistically significant.

\bgroup
\def\arraystretch{1.5}
\begin{table*}[]
\centering
\caption{Quantitative comparison of the segmentation performances with Turbolift and with reversed Turbolift (with postprocessing) for 4-fold cross-validation using the median±variance Dice, intersection over union (IoU), precision, sensitivity, and specificity (update data)}
\label{tab:metrics_rtl_4fold}
\begin{tabular}{ccccccc}
Dataset                                                             & Type & Dice        & IoU         & Precision   & Sensitivity & Specificity \\ \hline \hline
\multirow{2}{*}{CT}                                               & rTL      & 0.886±0.006 & 0.795±0.011 & 0.892±0.065 & 0.899±0.014 & \textbf{0.994±0.0}   \\
                                                                    & TL       & \textbf{0.895±0.006} & \textbf{0.81±0.008} &  \textbf{0.861±0.007} & \textbf{0.939±0.008} & 0.991±0.0   \\ \hline
                                                                    
\multirow{2}{*}{CBCT}                                               & rTL      & 0.879±0.004 & 0.784±0.008 & 0.846±0.008 & 0.929±0.005 & 0.99±0.0   \\
                                                                    & TL       & \textbf{0.895±0.006} & \textbf{0.81±0.008} &  \textbf{0.861±0.007} & \textbf{0.939±0.008} & \textbf{0.991±0.0}   \\ \hline
                                                                    
\multirow{2}{*}{\begin{tabular}[c]{@{}c@{}}CBCT\\ TST\end{tabular}} & rTL        & 0.882±0.012 & 0.79±0.015 & 0.875±0.009 & 0.913±0.019 & 0.993±0.0   \\
                                                                    & TL       & \textbf{0.905±0.007} & \textbf{0.826±0.01} & \textbf{0.882±0.005} & \textbf{0.921±0.011} & 0.993±0.0   \\ \hline
\end{tabular}
\end{table*}

\begin{figure}
    \centering
    \includegraphics[width=0.49\textwidth]{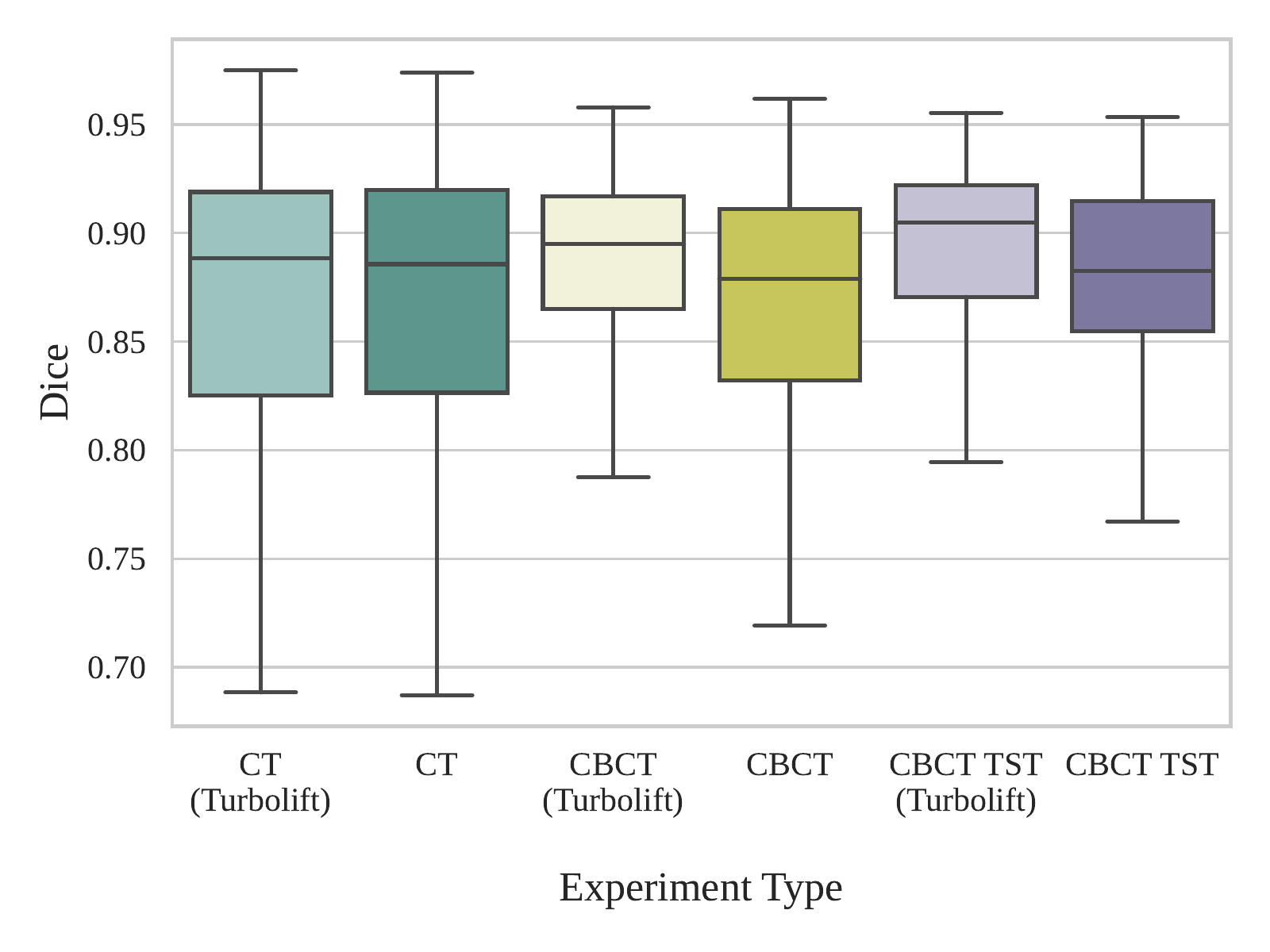}
    \caption{Dice scores for 4-fold cross-validation - comparing the Turbolift results against the results of reversed Turbolift}
    \label{fig:Dice_results_rvrsTL_4fold}
\end{figure}

\paragraph{Without CHAOS pre-training}
Pre-training the network model instead of initialising the network weights randomly can improve the performance of the model, as discussed in Sec.~\ref{sec:turbolift}. However, to evaluate whether this holds true for the current experimental scenario of Turbolift learning for the given CT, CBCT, and CBCT TST datasets, an additional set of trainings of the Turbolift was performed without using CHAOS pretrained models before training on the CT dataset. In this case, the weights were randomly initialised using a uniform distribution following the method described in~\citet{he2015delving}. After which, the model was trained following the Turbolift order: CT, CBCT, CBCT TST. The results are shown in Table.~\ref{tab:metrics_nochaos_4fold}, while the resultant Dice scores have been visualised with the help of box plots in Fig.~\ref{fig:Dice_results_noCHAOS_4fold}. The results show that for CBCT and CBCT TST, Turbolift with CHAOS pre-training performed better. In eight out of ten scenarios, the Turbolift with CHAOS pre-training resulted in higher scores with statistical significance than the Turbolift without CHAOS pre-training. For CBCT TST, both cases resulted in the same median specificity. And for CBCT TST dataset, the improvement observed in terms of sensitivity with the CHAOS pre-training was not statistically significant. From all these results, it can be said that the CHAOS pre-training helped Turbolift improve the performance on the CBCT and CBCT TST datasets. On the CT dataset, however, the trend is totally different. In all five metrics, a drop in scores can be seen with CHAOS pre-training - four of which (except sensitivity) are also statistically significant. This contradicts the earlier findings~\citep{bengio2017deep,chatterjee2022classification}. One possible answer that it might be because of the fact that the CT (as well as CBCT and CBCT TST) dataset was of animal abdomens, while the CHAOS dataset was of human beings. Moreover, the CT dataset had slices with contrast agent, which was not present in the CHAOS dataset. These two factors might have negatively contributed to the result of the model, because the CT training was the first stage of Turbolift. After the model is already got trained on CT after CHAOS pre-training, the task became easier for CBCT and CBCT TST trainings. This still does not answer the question of why CBCT and CBCT TST resulted in higher scores with CHAOS pre-training when the initial stage of Turbolift (i.e. CT) resulted in lower scores. This demands further experiments and analyses.  

\bgroup
\def\arraystretch{1.5}
\begin{table*}[]
\centering
\caption{Quantitative comparison of the segmentation performances with and without CHAOS pre-training preceding Turbolift (with preprocessing) for 4-fold cross-validation using the median±variance Dice, intersection over union (IoU), precision, sensitivity, and specificity}
\label{tab:metrics_nochaos_4fold}
\begin{tabular}{ccccccc}
Dataset                                                             & With CHAOS & Dice        & IoU         & Precision   & Sensitivity & Specificity \\ \hline \hline
\multirow{2}{*}{CT}                                             & No    & \textbf{0.896±0.006} & \textbf{0.811±0.011} & \textbf{0.879±0.008} & \textbf{0.939±0.009} &  \textbf{0.991±0.0}   \\ 
                                                                  & Yes    & 0.888±0.006 & 0.799±0.01 & 0.861±0.009 & 0.937±0.007 & 0.937±0.007   \\ \hline
\multirow{2}{*}{CBCT}                                               & No      & 0.887±0.006 & 0.798±0.009 & 0.84±0.007 & 0.936±0.008 & 0.989±0.0   \\
                                                                    & Yes       & \textbf{0.895±0.006} & \textbf{0.81±0.008} &  \textbf{0.861±0.007} & \textbf{0.939±0.008} & \textbf{0.991±0.0}   \\ \hline
                                                                    
\multirow{2}{*}{\begin{tabular}[c]{@{}c@{}}CBCT\\ TST\end{tabular}} & No        & 0.882±0.012 & 0.79±0.015 & 0.875±0.009 & 0.913±0.019 & 0.993±0.0   \\
                                                                    & Yes       & \textbf{0.905±0.007} & \textbf{0.826±0.01} & \textbf{0.882±0.005} & \textbf{0.921±0.011} & 0.993±0.0   \\ \hline
\end{tabular}
\end{table*}

\begin{figure}
    \centering
    \includegraphics[width=0.49\textwidth]{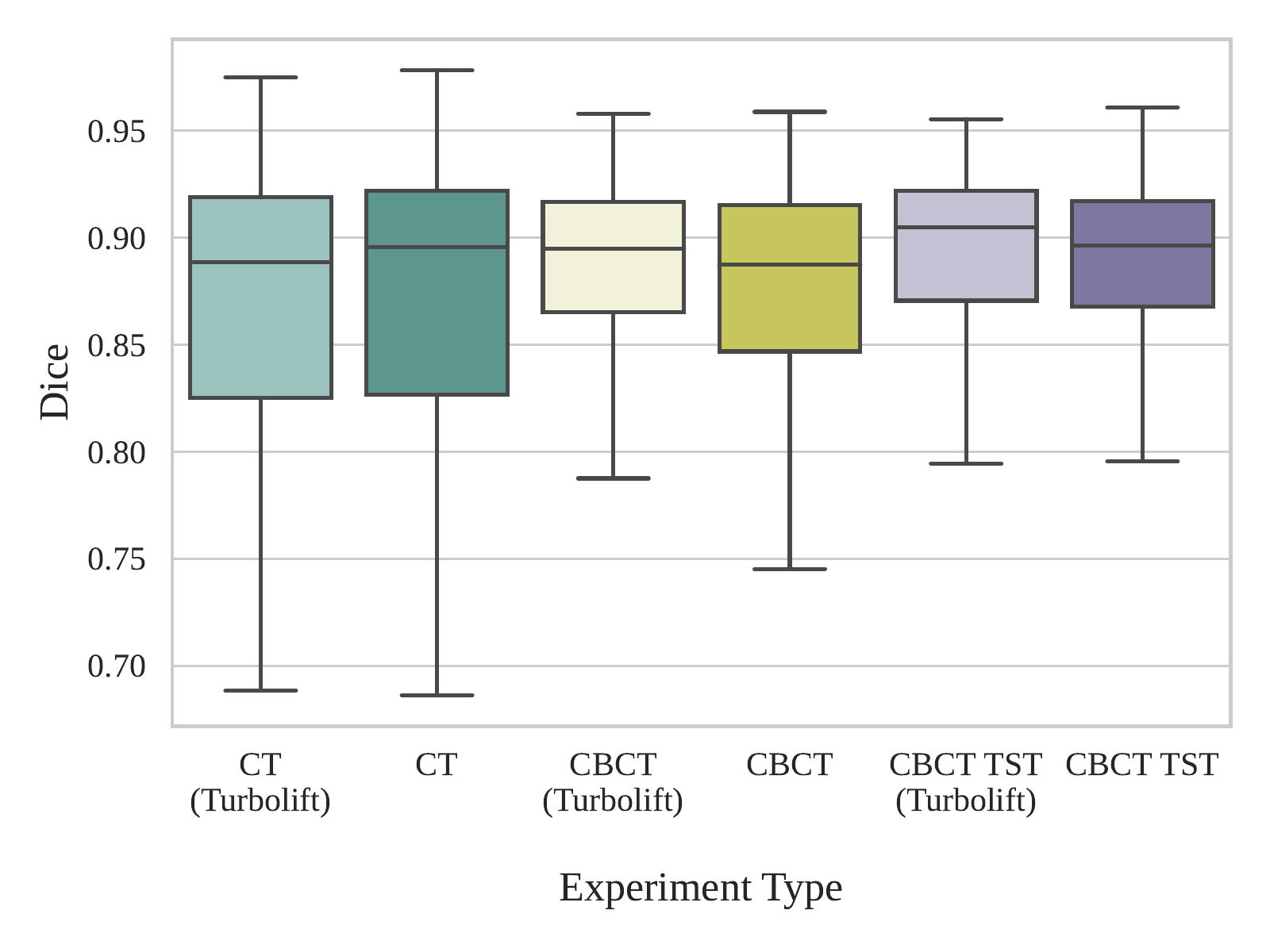}
    \caption{Dice scores for 4-fold cross-validation - comparing the Turbolift results against the results of obtained without pre-training Turbolift using the CHAOS dataset}
    \label{fig:Dice_results_noCHAOS_4fold}
\end{figure}

\paragraph{Testing with Artefacts}
The trainings were done on the datasets that include truncation artefacts and noise, but not the artefacts caused by the materials used for embolisation. The different embolisation materials caused these artefacts to appear less severe in some datasets, unlike in others. To evaluate the robustness of the model, the trained models were tested on data with such artefacts. This artefact was seen in only a few slices.
Considering the limited amount of data, these were not included in the training and were only used in this in-depth analysis of the Turbolift. The predicted masks for slices with severe artefact are shown in Fig.~\ref{fig:a1508}. The median Dice and IoU values (along with their variances) are given in Table~\ref{tab:metrics_art_4fold}. It can be seen that even though the models faced some difficulties, they managed to segment the decently, as indicated by the metric values. Moreover, these same slices with artefacts were also supplied to the models trained without Turbolift learning (the models shown in Table~\ref{tab:metrics_main_4fold}). In this case also, it can be seen that the Turbolift learning improves the robustness of the models against artefacts, resulting in better Dice and IoU scores.

\begin{figure}
    \centering
    \includegraphics[width=0.48\textwidth]{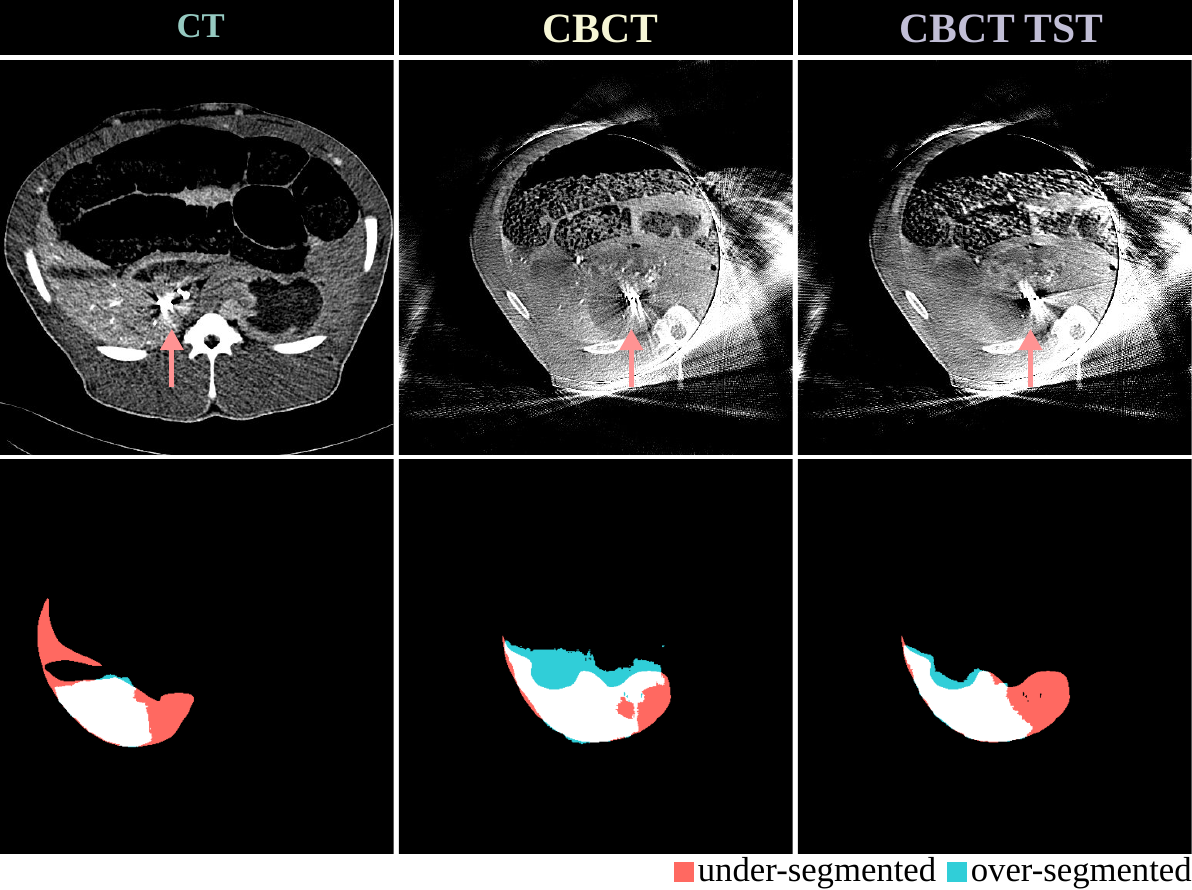}
    \caption{Example of strong artefacts caused by the embolisation material - red arrows are pointing to the source of the artefacts and the results of Turbolift learning while working with those slices.}
    \label{fig:a1508}
\end{figure}

\bgroup
\def\arraystretch{1.5}
\begin{table}[]
\centering
\caption{Quantitative estimation of the segmentation performances of slices with embolisation artefacts with Turbolift (with preprocessing) for 4-fold cross-validation using the median±variance Dice and intersection over union (IoU)}
\label{tab:metrics_art_4fold}
\begin{tabular}{cccc}
Dataset & Turbolift & Dice & IoU         \\ \hline \hline
CT                                             & Yes/No & 0.698±0.006 &  0.536±0.01   \\ \hline

\multirow{2}{*}{CBCT}                                       & No & 0.654±0.091 & 0.488±0.057   \\ 
                                               
                                        & Yes & \textbf{0.697±0.098} & \textbf{0.535±0.063}   \\ \hline

\multirow{2}{*}{\begin{tabular}[c]{@{}c@{}}CBCT\\ TST\end{tabular}} & No &
   0.672±0.007 & 0.506±0.008  \\ 

& Yes &
    \textbf{0.745±0.002}  & \textbf{0.594±0.003}  \\ \hline
\end{tabular}
\end{table}

\section{Discussion}
This paper has shown quantitatively and qualitatively that the Turbolift learning improves the performance of the model when it comes to the CBCT segmentations (straightforward and TST). This chapter presents some nuances of the results. It is worth mentioning that the liver of the four animals used in this study was different in terms of shape and size - making the overall task of generalisability difficult. As expected, CT provided better results, even though it was the first stage of the Turbolift - meaning it received no additional pre-training apart from CHAOS, as it provides a clear delineation between the liver and the surrounding tissues compared to CBCT. However, the performance on the slices with less visible vessels was better - which could be due to the fact that most of the available volumes did not have enough vessel visibility as they were taken before contrast agent inflow or after its outflow. But this difference in performance was less in the 4-fold experiments than in the 6-fold ones. Another important observation that can be made from the results is the model mostly managed to keep the non-liver tissues out of the liver segment (as desired) with Turbolift learning, even though the liver and the surrounding tissues have similar attenuation coefficients - resulting in an overall better segmentation of the liver with Turbolift than the trainings which were performed directly on the given dataset without Turbolift. This is more pronounced in the 4-fold results than in the 6-fold results, as can be seen in Fig~\ref{fig:results_GB}. In CBCT volumes during which the contrast agent passes the organ, the embolised region can be clearly differentiated from the rest of the liver tissue but also the grey values are close to the surrounding non-liver tissues. Still, this region was always well segmented. Another interesting observation was regarding the gallbladder. This is an organ with distinct attenuation values compared to its surroundings as they did not have any contrast agent in them - they clearly appear darker than the neighbouring regions. But, the model without using Turbolift also segmented the gallbladder as part of the liver. The authors hypothesise that the model tries to predict a general shape of the liver - making it over-segment the gallbladder region as well. However, the models with Turbolift succeeded in segmenting it out. Moreover, this exclusion of gallbladder was better in 4-fold than in the 6-fold results. While comparing CBCT and CBCT TST results with Turbolift, it can be seen that the CBCT TST resulted in the exclusion of the gallbladder better than CBCT. It is important to exclude the gallbladder from the segmentations as it can be used as a marker for image registration. For this current research, it is important that the model should not treat the vessel branching differently - they should be considered as parts of the liver - which the models with and without Turbolift managed to achieve for both CBCT and CBCT TST, unlike the CT results. A general comparison of the results revealed that the inclusions of the vessels were better in CBCT than in CT. In the case of CBCT TST, although only the first coefficient was used in this research for liver segmentation and the other ones are used for the modelling of the dynamics of the organs still, the regions with vessels are visible. Overall, the middle slices within the volumes resulted in better segmentations than the peripheral slices for all three types of images. Finally, it was observed that the CBCT resulted in over-segmentations, while CT and CBCT TST resulted in under-segmentations. It is worth mentioning that it is better to have under-segmentations than over-segmentations for the CT as SVD is applied on CT volumes, and since this is used to model CBCT liver voxels, it is quintessential to avoid non-liver voxels. 4-fold also resulted in less over-segmentations on the CT dataset than 6-fold. In every experimented scenario, it was observed (and as also discussed here) that the 4-fold experiments resulted in better performance than 6-fold - which can be attributed to the fact that the 4-fold experiments had three animals for training in each of its folds, while 6-fold had two. This also indicates that increasing the number of animals in the training set might also improve the overall performance of the model on every dataset.  

The in-depth analyses performed in this paper (Sec.~\ref{sec:indepth}) provided more insights into the method and the results. The first of such analyses, the flipped Turbolift (fTL) experiment, demonstrated that the performing CBCT training before training on CBCT TST improves the segmentation performance on the CBCT TST dataset, while training first on CBCT TST before training on CBCT did not help CBCT segmentations. This not only confirms the selected order of Turbolift (Sec.~\ref{sec:turbolift}), but also shows that the problem of the lack of a large CBCT TST dataset was successfully mitigated by training on CBCT before CBCT TST training. The revered Turbolift (rTL) experiment further confirms the chosen training order of Turbolift. These results show that the CT training does help CBCT, but CBCT does not help CT. The results also shows that the CT results do not improve neither with CHAOS pre-training, nor with additional two stages of pre-training (CBCT TST and TST). The authors hypothesise that due to the differences in image contrast in CT, it is difficult for the network to segment it perfectly - the training order of the original Turbolift narrows down the task for the network, while the order for the reversed Turbolift expands it. One more noteworthy difference between the TL and both rTL and fTL is the size of the datasets. In the chosen order of TL, the largest (in terms of the number of volumes) dataset was used at the first stage and the smallest at the last (discussed in Sec.~\ref{sec:perfscan}). rTL and fTL do not follow this order in terms of the size of the datasets - which could have also caused them to have poorer performance. As a final note regarding the CBCT TST dataset used in this research, only the first coefficient for each animal was used for the CBCT TST segmentations without computing time-resolved volumes, even though it is possible to obtain more data by computing volumes out of these coefficients. This was because the first coefficient resembles the most to the mask sweeps of the straightforward reconstruction and this satisfies the requirements of the use-case scenario presented in Fig.~\ref{fig:usage}. Furthermore, as discussed earlier, the best segmentation performance on CBCT was obtained for the mask sweeps (i.e. without contrast agent passing through the organ), and this also supports the selection of only the first coefficients from the CBCT TST datasets. Due to their resemblance with the first coefficient of CBCT TST, the masks obtained from the CBCT TST dataset, which resulted in better scores than CBCT, will also be applicable to CBCT dataset.  

The Turbolift learning attempts to combat the problem of small dataset by learning in different stages. The focus of this research has been on segmenting different types of perfusion images. Nonetheless, this kind of multi-stage training strategy can also be used for other tasks, like classification or object detection, where there are relations between the different stages - like the relation in the case of this research among CT, CBCT and CBCT TST.


\section{Conclusion and Future Work}
This research presents Turbolift learning, which trains and tests a modified version of the multi-scale Attention UNet model on different types of perfusion data one after the other, where the earlier training stages act as pre-training for the subsequent ones, while also generating the required results. The final stage of this experiment (CBCT TST), resulted in a Dice of 0.905±0.007 in a 4-fold cross-validation setup - improving it from 0.882±012, achieving statistically significant improvement over the model, which was only trained on the CBCT TST dataset without Turbolift. In different experiments and evaluation metrics, Turbolift learning resulted in statistically significant improvements between $0.8\%$ and $5.47\%$ over the respective baseline. Moreover, the Turbolift learning aided the model in becoming more robust against artefacts originating from the embolisation material. Further evaluations will be performed to evaluate if the model suffers from catastrophic forgetting, while the directions towards continual learning will also be explored. Furthermore, trainings will also be performed by including the slices- with artefacts to make them more robust against such artefacts.

\section*{Acknowledgement}
This work was in part conducted within the context of the International Graduate School MEMoRIAL at Otto von Guericke University (OVGU) Magdeburg, Germany, kindly supported by the European Structural and Investment Funds (ESF) under the programme "Sachsen-Anhalt WISSENSCHAFT Internationalisierung" (project no. ZS/2016/08/80646) and by the German Federal Ministry of Education and Research within the Research Campus STIMULATE (grant no. 13GW0473A and 13GW0473B).

\bibliography{mybibfile}

\end{document}